\documentclass[aps,prx,showkeys,superscriptaddress,reprint,floatfix]{revtex4-2}

\usepackage[utf8]{inputenc}
\usepackage[T1]{fontenc}
\usepackage{amsmath}
\usepackage{amssymb}
\usepackage{amsfonts}
\usepackage{upgreek}
\usepackage{mathtools}
\usepackage{siunitx}
\usepackage{float}
\usepackage{graphicx}
\usepackage{tabularx}
\usepackage[labelfont=bf]{caption}

\usepackage[hidelinks]{hyperref}
\usepackage[normalem]{ulem}
\usepackage{xcolor}

\usepackage{soul}
\usepackage{braket}
\usepackage[ruled, noline]{algorithm2e}

\begin{document}

\title{Multiscale Embedding for Quantum Computing}

\author{Leah P. Weisburn}
\thanks{Denotes Equal Contribution}
\affiliation{Department of Chemistry, Massachusetts Institute of Technology, Cambridge, MA 02139, USA}

\author{Minsik Cho}
\thanks{Denotes Equal Contribution}
\affiliation{Department of Chemistry, Massachusetts Institute of Technology, Cambridge, MA 02139, USA}

\author{Moritz Bensberg}
\affiliation{ETH Zurich, Department of Chemistry and Applied Biosciences, Vladimir-Prelog-Weg 2, 8093 Zurich, Switzerland.}

\author{Oinam Romesh Meitei}
\affiliation{Department of Chemistry, Massachusetts Institute of Technology, Cambridge, MA 02139, USA}

\author{Markus Reiher}
\affiliation{ETH Zurich, Department of Chemistry and Applied Biosciences, Vladimir-Prelog-Weg 2, 8093 Zurich, Switzerland.}

\author{Troy Van Voorhis}
\email{tvan@mit.edu}
\affiliation{Department of Chemistry, Massachusetts Institute of Technology, Cambridge, MA 02139, USA}

\begin{abstract}

We present a novel multi-scale embedding scheme that links conventional QM/MM embedding and bootstrap embedding (BE) to allow simulations of large chemical systems on limited quantum devices. We also propose a mixed-basis BE scheme that facilitates BE calculations on extended systems using classical computers with limited memory resources. Benchmark data suggest the combination of these two strategies as a robust path in attaining the correlation energies of large realistic systems, combining the proven accuracy of BE with chemical and biological systems of interest in a lower computational cost method. Due to the flexible tunability of the resource requirements and systematic fragment construction, future developments in the realization of quantum computers naturally offer improved accuracy for multi-scale BE calculations.

\end{abstract}

\keywords{Hybrid Quantum Mechanics/Molecular Mechanics (QM/MM), Electrostatic Embedding, Quantum Computing, Correlated Electronic Structure, Drug Binding Energy}

\maketitle

\section{Introduction}\label{sec:intro}
Recent advances in quantum computing show potential for the efficient and accurate simulation of chemical systems on quantum devices. \cite{Bauer-QAforQC2020,McArdle_QCQChemReview-2020,Chan-QC2024}
The exponential growth in the multi-body Hilbert space with system size limits the application of highly-accurate methods on classical computers, especially for large systems of biological interest. 
Phenomena like drug binding and charge transfer processes are treated most accurately with electronic structure methods, yet the physical systems of interest are too large for most traditional quantum chemistry on classical computers.\cite{Baiardi_QCForMolecularBio-2023}
The inherently quantum mechanical nature of quantum computers, coupled with clever algorithms taking advantage of their architecture, suggests a path towards accurate modeling of these processes and systems. 
With enough high-fidelity qubits, methods like quantum phase estimation (QPE) can directly simulate extended systems with controllable accuracy.\cite{Kitaev_originalQPE1995,Abrams_QPEEigenvalues-1999,Reiher_QPERxnMechanisms2017}

Existing hardware, however, only allows for the simulation of very small systems. 
Current state-of-the-art quantum computers are limited to $O(10)$ logical qubits\cite{google_errorcorrecting, coldatom_errorcorrecting, Reichardt2024, Paetznick2024} and shallow circuit depth, inherently limiting the chemical calculations which can be performed and scientific questions which can be answered. 
These noisy intermediate-scale quantum (NISQ) devices have shown promise in various scenarios, with a non-exhaustive list of computational applications including simulating chemical Hamiltonians\cite{Google-HFonQC2020, vqe_ucc}, exploring topological order\cite{Satzinger_2021}, and modeling electron dynamics\cite{Babbush_2023}.    
The application of NISQ computers to extended biological systems remains a non-trivial target.
    
Various techniques have been proposed and developed to facilitate classical and quantum computing calculations describing large systems. Broadly speaking, many popular methods, and also the one introduced in this work, reduce a physical system to a small fragment or space that is directly simulated with the quantum computer. We loosely separate such overlapping methodologies into several categories of approach. 
    
Most generally, active space methods choose a subset of orbitals of interest within a system and perform the calculation within this space. Approaches like CAS-SCF\cite{Roos_original-CASSCF-1987,Werner_originalMCSCF-1985,Knowles_originalMCSCF-1985} and its derivatives in traditional electronic structure theory\cite{Pandharkar_LAS-SI-2022}  and LAS-UCC or quantum CAS-SCF in quantum computing are used to describe molecular dissociations, reactions and more.\cite{Tilly_QCASSCF-and-Embedding-2021,deGraciaTrivino-CAS-NISQ-OrbitalOpt2023,Gujarati_QC-RxnsWEmbedding2023,Otten_LAS-UCC-2022, D_Cunha_DILASUCC-code-2024, Battaglia_AS-QC-2024,Izsak_QM-Multilayer-2022} 
    
Several approaches employ downfolding, which we characterize as a global method which chooses a model Hamiltonian to represent the full system.\cite{Downfolding-Book} Methods like driven similarity renormalization group, quantum unitary downfolding, and derivatives of coupled cluster like SES-CC and DUCC use this idea to study bond breaking of molecules, Ising models, and more.\cite{Evangelista_DrivenSimilarityRenormalizationGroup-2014,Bauman_CC-DoubleCommuterTerms-2022,Metcalf_VQE-DUCC2020,Kowalski_DUCCAccuracies-2024,Kowalski_DimensionalityRedCC-2021,Kowalski_QFLOW-2023,Romanova_DynamicalDownfoldingLocalQS-2023,Huang-downfolding-2023,Misiewicz_PQE-Implementation-2024}
    
Finally, we distinguish embedding methodologies as local approaches which break a system into fragments which are reassembled to represent the entire system. A non-exhaustive list of embedding methods and divide-and-conquer approaches\cite{Yang_DivideAndConquer-1992} in traditional quantum chemistry include ONIOM,\cite{HONIG_first-ONIOM-1971,Warshel_second-ONIOM-1976,Chung_ONIOM-Review-2015} DFT-based embedding techniques \cite{Jacob_SubsystemDFT-2014,Jacob_SubsystemDFT2-2024,Libisch_WF-in-DFT-2014, Manby_DFT-Embedding-2012,Lee_Projection-WV-in-DFT-2019, Rossmannek_QuantumEmbeddingonQC-2023}, Huzinaga embedding\cite{H_gely_Huzinaga-2016,Chulhai_GoodpasterHuzinaga-2017,Bensberg_BasisSetAdaptation-Projection-2019}, density matrix embedding theory (DMET)\cite{Knizia_DMET-2012}, bootstrap embedding (BE)\cite{dmet-boot-1,dmet-boot-2,be-atom,be-ccsd,be-mol,be-ube}, and others\cite{Wesolowski_FrozenDensity-2015}. These methods have already been extended to quantum computers, including DMFT\cite{Kreula_QC-DMFT-Exp-2016}, DMET\cite{Kawashima_DMET-on-QC-2021}, projection embedding\cite{Rossmannek_QuantumEmbeddingonQC-2023}, QDET\cite{Sheng_GW-QDET2022}, and more\cite{Ma_Galli-NISQ-QC-2020, Ma_MultiscaleQAforQC-2023} for Hubbard models, hydrogen rings, and other small chemical systems.

Algorithms like QPE can provide a substantial speedup over classical algorithms for computing the eigenvalues of chemical Hamiltonians. This could lead to dramatic advances in the quantum simulation of large systems. While there is discussion as to the practical advantages and limitations of quantum computers in quantum chemistry, the new technology inspires innovation and further investigation.\cite{Lee-ExpQuantumAdvantage2023} Here, we work to show that a multiscale approach involving classical embedding, quantum embedding, and a hierarchy of quantum calculations facilitates the rapid simulation of thousands of atoms in such a way that near-term quantum computing could be applied for extended systems. To construct small Hamiltonians that can be solved on near-term quantum devices, we for the first time couple traditional hybrid quantum mechanics-in-molecular mechanics embedding (QM/MM) with a bootstrap embedding (BE) quantum mechanics-in-quantum mechanics (QM/QM) approach for the QM region. This BE method reduces the QM region into a series of fragments such that the entire system, not just a small active space, is treated evenly with a correlated method like CCSD on classical computers or QPE on quantum computers. We also introduce a new mixed-basis embedding method that further allows for BE to be applied to large QM regions with dense atomic orbital basis sets, better posed to address large biological systems. As the development of quantum computers continues, this method allows the technology to be directly applied to chemical systems of interest, beyond small toy models, with a tunable number of orbitals and thus qubits.

We note that the successful application of quantum computers for computational chemistry requires various considerations beyond tunable control over the number of qubits. In particular, for workflows that utilize QPE as the eigensolver, state preparation is a key concern, as a guiding state with enough overlap must be generated and provided. To this end, previous work has shown that mean-field ground states serve as robust guiding states for BE Hamiltonians,\cite{paper2} which reduces the required quantum circuit depth for robust phase estimation.\cite{robust_phase_estimation,paper4}

Still, to obtain sensible results with QPE, we require quantum devices which satisfy qubit size and circuit depth requirements. A related work on resource estimation shows that chemical Hamiltonians of moderate size at the upper end of large FCI calculations ($\sim$ 16 spatial orbitals) require about $10^6 - 10^7$ two-qubit gates.\cite{umbrella} The study predicts, based on device roadmaps, that a gate count of about $10^5$ would be attainable within the next 3 years. The work claims that the gap is modest enough with advances in error correction to anticipate QPE-based workflow for quantum embedding Hamiltonians like BE in the near future.\cite{umbrella} In this work, we propose strategies that mainly focus on two aspects of quantum computing for chemical applications: (i) number of available qubits and (ii) the interplay between classical high-performance computing and quantum computers.

In this paper, we introduce in Section \ref{sec:qm-qm-mm} a QM/QM/MM method that couples QM/MM with BE to address the quantum mechanical region of a large system at a correlated wavefunction level. In Section \ref{sec:mixed-basis}, we next develop a mixed-basis BE method that allows for the calculation of increasingly large systems on classical computers. 
In Section \ref{sec:comp_details}, we illustrate the practical details including how we automatically choose the QM regions consistently for an ensemble of large biochemical structures.    
We successfully demonstrate the mixed-basis QM/QM/MM method for biological systems, specifically the well-studied protein-ligand complex of myeloid cell leukemia 1 and the inhibitor 19G\cite{Friberg2012} in Section \ref{sec:results}. We conclude with a discussion of our results in Section \ref{sec:discussions} and the application of the methods towards NISQ computers in Section \ref{sec:conclusion}.

\section{Theory}\label{sec:theory}

\subsection{QM/QM/MM} \label{sec:qm-qm-mm}
In this work, we build upon the classical QM/MM framework, which reduces the cost of modeling large chemical and biological systems.\cite{Chung_ONIOM-Review-2015,HONIG_first-ONIOM-1971,Warshel_second-ONIOM-1976} Within the electrostatic QM/MM methodology, a system is decomposed into a small region of interest treated with a QM method surrounded by an environment described at the MM level. This combination of regimes, in our case taken along steps in a molecular dynamics (MD) trajectory for a system, dramatically reduces the cost of the calculation while maintaining accuracy for the local chemical changes of interest (i.e. bond binding and breaking) within the QM region compared to a full QM treatment. The details by which we perform this QM/MM embedding and choose the QM region for large, biological systems are described in Section \ref{subsec:qm-selection}.

\begin{figure}[tb]
    \centering
    \includegraphics[width=0.9\linewidth]{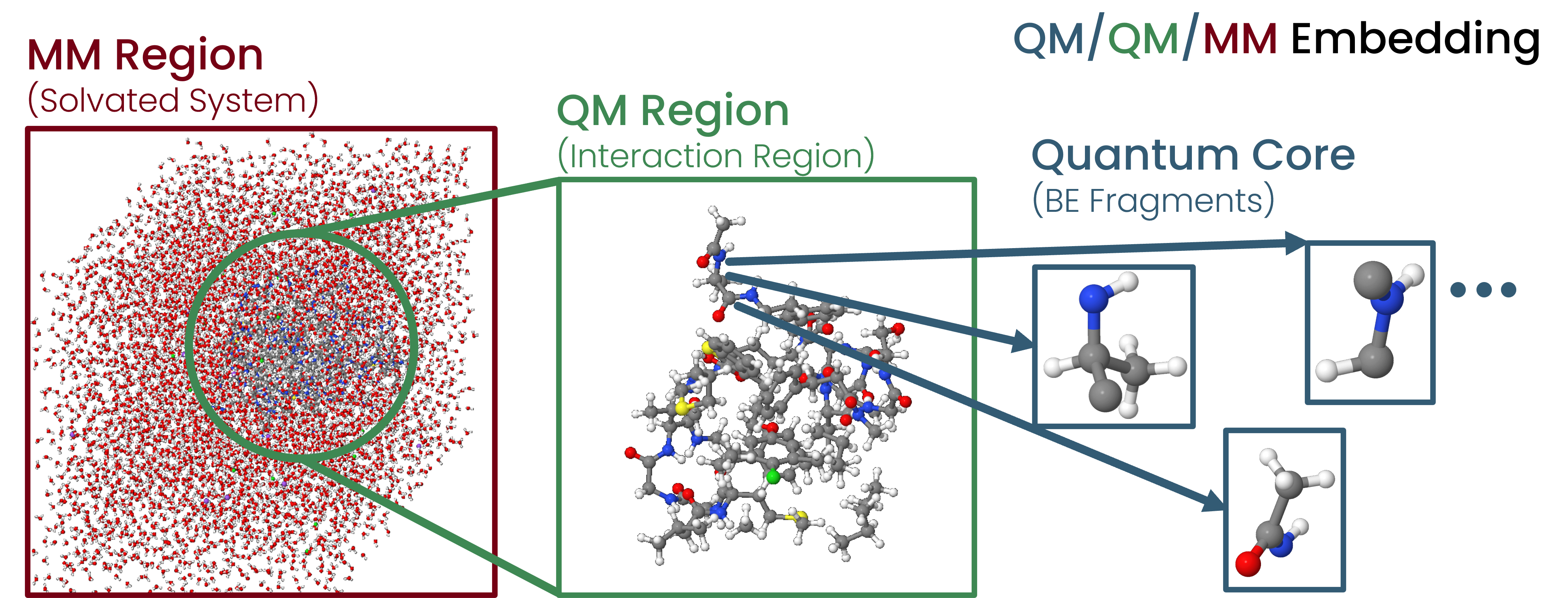}
    \caption{Diagram of the multi-layer QM/QM/MM embedding scheme}
    \label{fig:qmqmmm}
\end{figure}

Even with the traditional division of biological systems into QM and MM regions, the size of the quantum mechanical region necessary to capture the significant quantum effects for a given system tends to be much larger than the resource available on cutting-edge near-term quantum devices.\cite{Katabarwa_2024} Therefore, we employ BE to reduce the resources required to simulate the QM regions. In classical computing, BE has exhibited robust recovery of local correlation effects,\cite{be-atom, be-ccsd, be-mol, dmet-boot-1, dmet-boot-2, be-ube, k-be} and thus provides a powerful tool we may build upon for these purposes.\cite{Liu-be-qc_2023} Extending both QM/MM methods and BE, we propose a multilayer embedding scheme shown in Figure \ref{fig:qmqmmm}. Within the classical \textit{MM regions}, we identify the chemically crucial portions as \textit{QM regions} for quantum mechanical descriptions. We then build the BE fragments, which we can either evaluate classically with a method like coupled-cluster or load on limited near-term quantum devices as the \textit{quantum core} of the calculations. By linking conventional QM/MM embedding and BE, we propose a novel extension to these embedding methods that expands the size of the systems one can model on limited hardware while retaining an accurate description of the quantum effects.

\subsubsection{Bootstrap Embedding}\label{subsec:BE}

Before explaining the multi-layered QM/QM/MM embedding in detail, we briefly introduce the essential concepts of bootstrap embedding (BE). We direct readers to previous publications for a more detailed discussion of the method.\cite{be-atom, be-ccsd, be-mol, dmet-boot-1, dmet-boot-2, be-ube, k-be, Liu-be-qc_2023} 

BE, a derivative of density matrix embedding theory (DMET)\cite{Knizia_DMET-2012}, reduces a chemical system into overlapping fragments, each of which can be treated evenly at a high level with accurate wavefunction methods and reassembled to describe the full system. BE calculations start with a Hartree-Fock (HF) wavefunction, our reference mean-field expression, of the quantum mechanical system of interest. We begin with the following second-quantized Hamiltonian that describes the chemical system.

\begin{equation}
    \Hat{H}_\text{QM} = \sum_{\mu \nu}^{\text{N}_\text{basis}} h_{\mu \nu} \hat{c}_\mu^\dagger \hat{c}_\nu + \sum_{\mu\nu\lambda\sigma} ^{\text{N}_\text{basis}} V_{\mu\nu\lambda\sigma} \hat{c}_\mu^\dagger \hat{c}_\lambda^\dagger \hat{c}_\sigma \hat{c}_\nu
\end{equation}

Here, $\mu,\nu,\lambda$, and $\sigma$ denote the working basis or atomic orbitals (AOs); $\hat{c}_{\mu}^{\dagger}$ and $\hat{c}_{\mu}$ are the creation and annihilation operators for orbital $\mu$; $h$ are the one electron integrals; and $V$ are the two electron integrals.

The mean-field wavefunction $\ket{\Psi_0}$ of the Hamiltonian $\Hat{H}_\text{QM}$ serves as the trial state used in the embedding procedure. After localizing the HF orbitals, we choose a subset of localized \textit{fragment} orbitals to form $\boldsymbol{F}$. Typically, we choose the fragment $\boldsymbol{F}$ to contain a small fraction of the orbitals comprising the full system.

Utilizing the Schmidt Decomposition, we then construct the entangled bath orbitals for the given fragment. We build an exactly reduced wavefunction around fragment $F$, with fragment orbitals $\ket{f^F}$ and entangled bath orbitals $\ket{b^F}$. This procedure also results in unentangled environment orbitals $\ket{e^F}$, which span the rest of the space. This reduced wavefunction in Equation \ref{eq:be_wfn} has a dimension of at most $2 N_{\text{frag}}^F$, where $N_\text{frag}^F$ is the number of fragment orbitals in the fragment $F$, no matter the size of the full system. $\Lambda^F_k$ are the Schmidt coefficients associated with each state.
\begin{align}
  \ket{\Psi_0} &= \sum_{k=1}^{N_{\text{frag}}^{\boldsymbol{F}}} \Lambda_{k}^{F} \left[ \ket{f_{k}^F} \otimes \ket{b_{k}^F} \right]  \otimes \ket{e^F} \label{eq:be_wfn}
\end{align}

The fragment and entangled bath orbitals are used to generate a reduced Hamiltonian representing the system, while the unentangled environment is treated simply at a mean-field level as an effective potential. Because of the greatly reduced dimension of this new Hamiltonian, it can be solved with a high-level wavefunction method which would else be intractable for the full system. This procedure is reiterated for multiple fragments in BE to produce a set of reduced Hamiltonians from the same HF reference.

\begin{figure}[tb]
    \centering
    \includegraphics[width=0.75\linewidth]{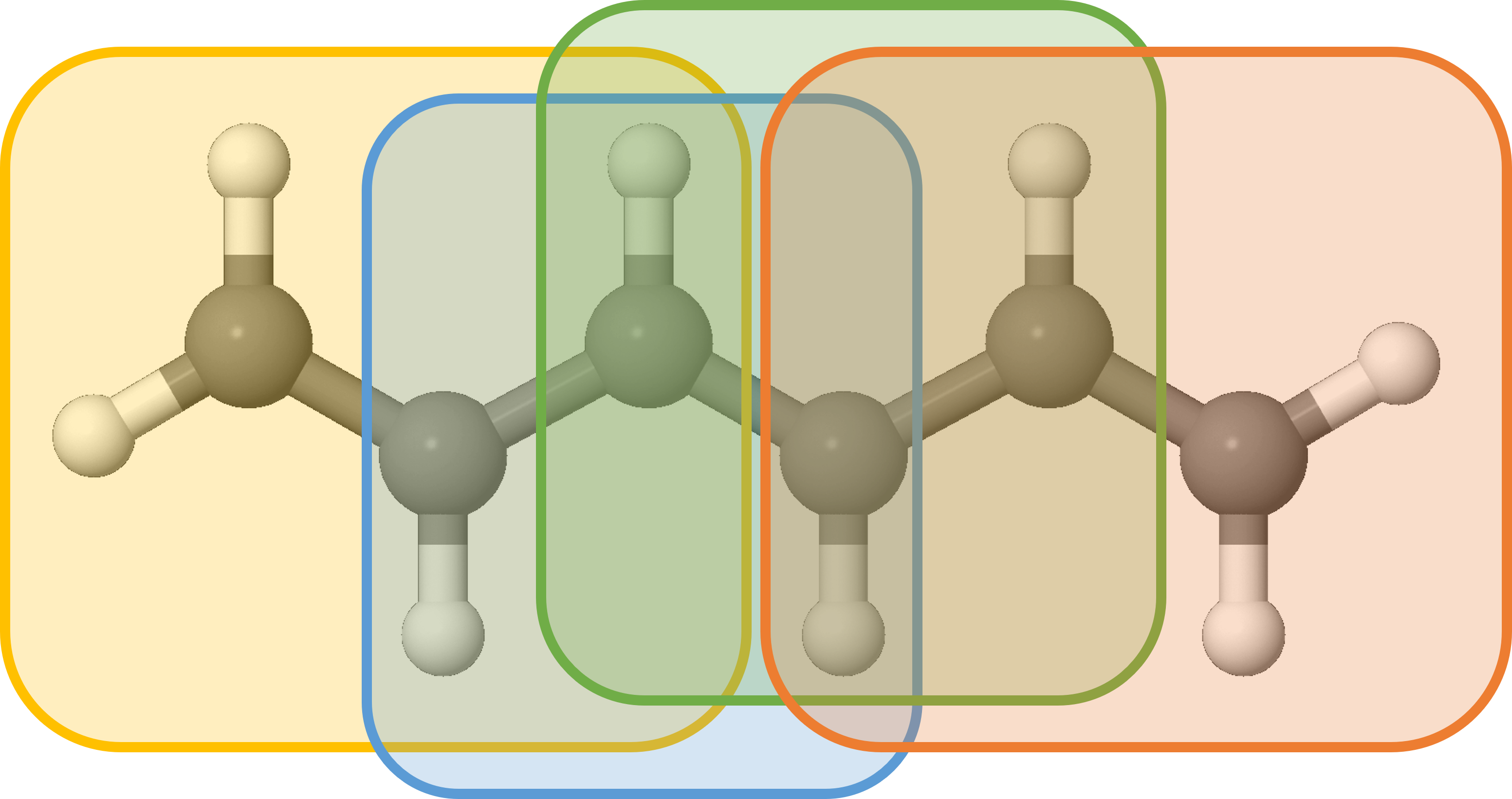}
    \caption{BE$(2)$ fragments of polyacetylene}
    \label{fig:be2polyac}
\end{figure}

BE is unique to other embedding methods in its construction of overlapping fragments based on atomic connectivity. In BE$(n)$, each fragment includes a center site atom and up to $(n-1)$-th nearest neighbors of the center site atom. For example, the BE$(1)$ fragment scheme builds fragments where each non-hydrogen atom forms a separate fragment. BE$(2)$ fragments include the nearest neighbors of the center site atoms so that the non-terminal atoms become members of multiple BE$(2)$ fragments. Figure \ref{fig:be2polyac} shows an example of BE$(2)$ fragments for a polyacetylene chain. A key assumption in BE is that each fragment calculation better represents the quantities (ex. 1- and 2-RDMs) corresponding to the \textit{center-site} atoms than that of the \text{edge-site} counterparts. Overlapping fragments reduce errors introduced by edge effects in other embedding methods.

We perform correlated calculations using high-level wavefunction methods such as CCSD, FCI, and DMRG within the formulated fragment space after constructing the interacting bath Hamiltonian.\cite{Knizia_DMET-2012} For self-consistency across multiple fragments, we optimize the global chemical potential which ensures a correct number of electrons in the chemical system, as discussed in previous works.\cite{be-atom, be-ccsd} We construct the energy, using a cumulant-based definition introduced in previous publication\cite{k-be} and in more detail in the Supporting Information, for the system by taking components of the reduced density matrix, cumulant, and Hamiltonian from each fragment calculation, careful to avoid overcounting.\cite{Nusspickel_Cumulant-2022, k-be, iao-paper} 

\subsubsection{Bootstrap Embedding with MM Charges}\label{subsec:be/qm/qm/mm}

We combine the BE quantum mechanics-in-quantum mechanics (QM/QM) method with conventional quantum mechanics-in-molecular mechanics (QM/MM) embedding to generate a QM/QM/MM embedding scheme suitable for describing large systems with tens of thousands of atoms. Within this scheme, we evaluate the QM region using BE and describe its interactions with its MM surroundings by employing electrostatic embedding.\cite{Warshel_second-ONIOM-1976} The MM charges influence the system Hamiltonian through the one-electron operators, in atomic units, below.

\begin{equation}
    \hat{h}_i^\text{QM-MM} = \hat{h}_i^\text{QM} - \sum_A^{N_\text{MM}} \frac{Q_A}{|r_i - R_A|}
\end{equation}
Here, $\Hat{h}_i^\text{QM}$ is the original one-body operator that includes the kinetic and nuclear interactions for the $i$-th electron. $A$ indexes through the $N_\text{MM}$ point charges, each assigned with $Q_A$ charge. In the second-quantized formalism, this is represented as an additional term $U$ that we evaluate using standard quantum chemistry packages. For this study, \texttt{pyscf}'s \texttt{qmmm} module incorporates the MM charges into the Hamiltonian of the QM region.\cite{pyscf}
\begin{multline} \label{eq:qmmm}
    \Hat{H}_\text{QMMM} = \sum_{\mu\nu}^{N_\text{basis}} (h_{\mu\nu}+U_{\mu\nu})\hat{c}_{\mu}^{\dagger}\hat{c}_{\nu} \\ + \frac{1}{2}\sum_{\mu\nu\lambda\sigma}^{N_\text{basis}} V_{\mu\nu\lambda\sigma}\hat{c}_\mu^\dagger \hat{c}_\lambda^\dagger \hat{c}_\sigma \hat{c}_\nu 
\end{multline}
\begin{equation}
    U_{\mu\nu} = -\sum_A \left<\mu \middle| \frac{Q_A}{|r-R_A|} \middle| \nu\right>
\end{equation}

We perform the subsequent BE steps using the electronic Hamiltonian in Equation \ref{eq:qmmm}. We obtain the Hartree-Fock trial wavefunction by performing a mean-field calculation for the dressed Hamiltonian. Fragmentation and bath construction steps are identical to a conventional BE calculation, simply using $\Hat{H}_\text{QMMM}$ instead of $\Hat{H}_\text{QM}$. This allows us to utilize the code previously developed for BE with relatively small modification.\cite{oimeitei/quemb:-2024-09-03}

In this study, we limit ourselves to \textit{one-shot} BE calculations, skipping the self-consistency and density matching loop performed in previous papers. While these self-consistency loops can trivially be applied in this procedure, in this instance, it is useful to limit the necessary quantum resource requirement in the context of near-term quantum computing applications. To assess the efficiency of the QM/QM/MM embedding, we use CCSD as a surrogate for Hamiltonian simulation on quantum computers, which will be explored in the near future.

\subsection{Mixed-Basis BE} \label{sec:mixed-basis}
While the BE method avoids the prohibitive cost of the high-level solver for the full system, it relies upon a system-wide mean field calculation and integral transforms for each of the fragments. For systems of interest with hundreds of atoms and large basis sets, this can become both memory and time prohibitive. 

\begin{figure}[tb]
    \centering
    \includegraphics[width=0.9\linewidth]{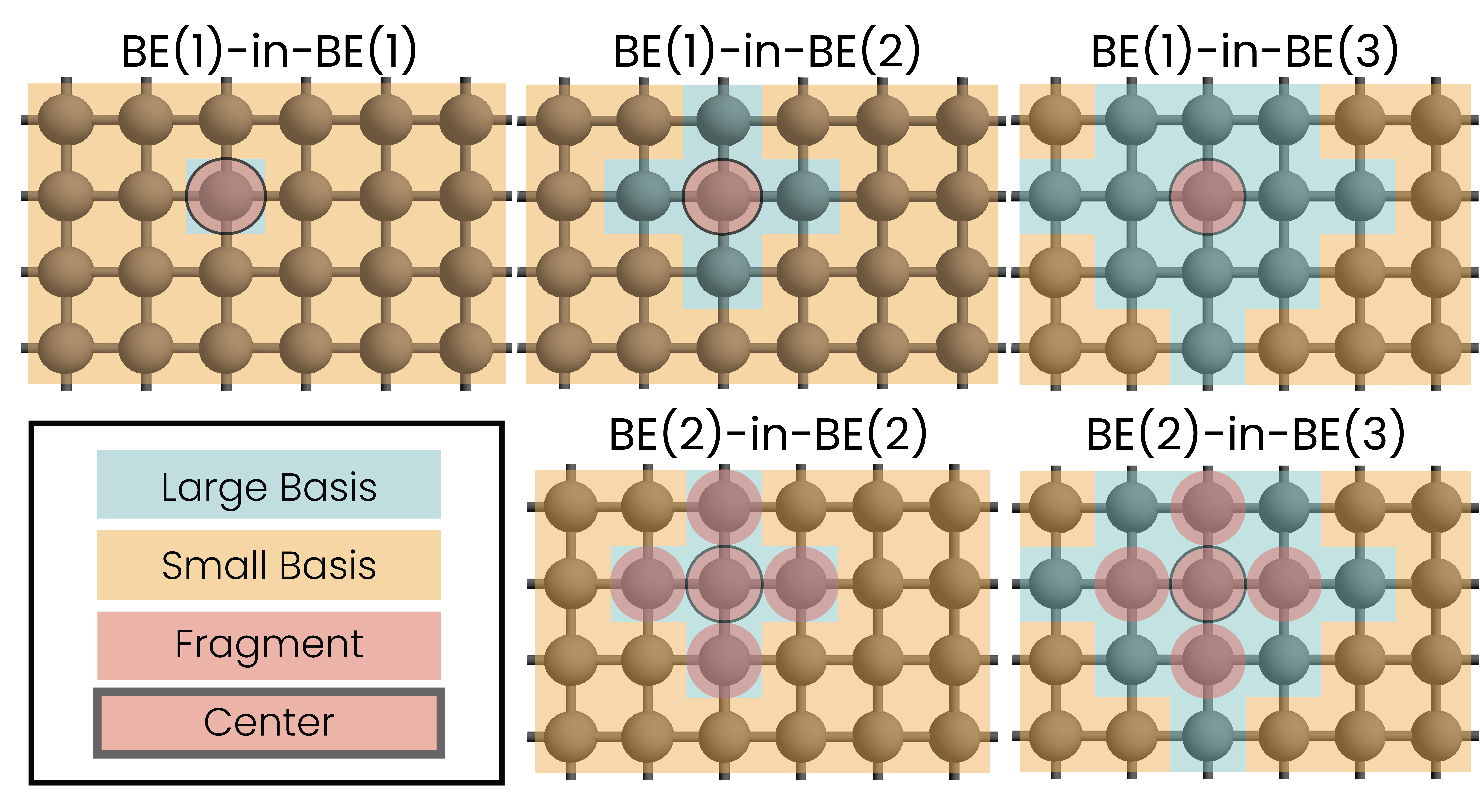}
    \caption{Mixed-basis fragmentation schemes. For a given outlined center atom, the BE$(n)$ fragment is highlighted red, the large basis set region of fragmentation prescription BE$(m)$ is highlighted green, and the rest of the system with small basis set is orange. The same fragmentation and mixed-basis scheme is followed for all atom centers in the system for a given chemical calculation.}
    \label{fig:mixed-basis-def}
\end{figure}

We introduce a mixed-basis embedding method which places a larger, denser basis near a given fragment and chooses a minimal basis set otherwise, aiming to dramatically reduce the cost of the calculation while minimally impacting the accuracy of the correlation energy.
For each of the BE$(n)$ fragments described in Section \ref{sec:qm-qm-mm}, we select a surrounding region of size prescribed by BE$(m)$ for which we choose a dense basis set.
We refer to this fragment and basis choice as BE$(n)$-in-BE$(m)$ embedding, as depicted in Figure \ref{fig:mixed-basis-def}.
The large basis region around a central atom for BE$(1)$-in-BE$(1)$ is thus only a given atom; in BE$(1)$-in-BE$(2)$, it is the atom and its nearest directly connected neighbors; for BE$(1)$-in-BE$(3)$, the region includes the two rings of nearest neighbors. 
With increasing $m$ to contain the full system within the BE$(m)$ region, the BE$(n)$-in-BE$(m)$ calculation converges to the BE$(n)$ prescription in the large basis set, and we indeed observe this convergence in the correlation energies for molecular systems.

It is important to note that mixed-basis approaches have been explored in depth atop other correlated methods over many years.\cite{Bauschlicher_MixedBasisSetHeBe-1982,Chesnut_LocallyDenseBasisSet-1989,Orozco_MixedBasisSetElectrostatics-1989,Hinton_AlInitioQMHWater-1992,DiLabio_BondDissocLocalDenseBasis-1998,DiLabio_LocallyDenseBasisMolec-1999,Reid_LocallyDenseBasisNMR-2013,Hegely_DualBasisSetEmbedding-2018,Cheng_MixedBasisPotFuncEmb-2017,FornaceUnknownTitle2015,Haldar_MultilayerIPEOMCCSD-2019,Mitra_PeriodicDMETCoMgO2022} In this work, we use the BE systematic black-box framework alongside the mixed-basis set idea to clearly define each basis set region. Unlike many other methods, which choose some subset of orbitals to be in a denser basis set than the others, we describe the entire system with the chosen large basis set across the set of calculations. This approximates and indeed converges to the absolute energy of the full system in the chosen large basis set.

This method requires separate mixed-basis HF calculations for each of the BE$(m)$ regions, as opposed to the single large basis HF calculation for conventional BE calculations. This prescription obviously creates a trade-off: the individual calculations are much faster in the mixed-basis, but more repetitions must be performed. Depending on the parameters of the problem, this trade-off may or may not be advantageous. We discuss the computational scaling more rigorously in Section \ref{subsec:mb_be_scaling}. For the large biological systems that we target in this work, performing many reduced-size HF calculations and the modified integral transformations for each proves less costly in time and memory than the full, non-reduced alternative.

\section{Computational Details} \label{sec:comp_details}
\subsection{Automatic QM Region Selection} \label{subsec:qm-selection}
When running large-scale high-throughput QM/MM or QM/QM/MM calculations for ensembles of large biological systems, manual division between the QM and MM regions becomes cumbersome. Here, we present an automatic procedure that identifies QM regions for a set of structures sampled from an MD trajectory. To generate a QM region, we introduce a variant of the automated QM region construction algorithm proposed in Reference \citenum{Brunken2021}. The original approach \cite{Brunken2021} generates a QM region such that the forces calculated with a QM/MM model for a selected atom matches the forces calculated for a very large reference QM region with the constraint of a maximum number of atoms $n^\mathrm{region}_\mathrm{max}$ in the final QM region. In our updated version of the algorithm, we perform this analysis for all ligand atoms, rather than for a single atom. Furthermore, we consider not only a single molecular structure to generate the QM region but a set of structures sampling the accessible configuration space at given conditions. This procedure results in a large QM region, surrounded by MM point charges, describing the biological system of interest at multiple points in the trajectory. We describe this procedure in detail in the Supporting Information.
\subsection{Biological Target System Preparation}
We investigate the small drug molecule 19G bound to the protein myeloid cell leukemia 1 (MCL1)\cite{Friberg2012} in this study. The task of MCL1 is to regulate apoptosis, and its dysregulation is found in connection to various cancers. As such, multiple binders, such as 19G, inhibiting its activity have been developed\cite{Wang2021}.

To obtain example structures that represent a typical biological system, we take snapshots from an MCL1-19G classical MM-MD trajectory\cite{paper1}. Because the ligand and its interaction with the protein are crucial to calculating the binding free energies in protein-drug complexes, we focus on the 19G ligand and its surroundings for the QM region selection.

We calculate the forces necessary for the automatic QM region selection procedure with the semi-empirical tight binding method GFN2-xTB \cite{Bannwarth2019}. As discussed in Section \ref{subsec:qm-selection}, we extract a total of 100 snapshots equally spaced in time from the MM trajectory of the MCL1-19G protein--ligand complex of Reference~\citenum{paper1}. The parameters discussed in the Supporting Information for the QM region selection are chosen as follows:
We set the maximum size of the reference QM region $n^\mathrm{ref}_\mathrm{max}$ to $500$ atoms, the maximum size of the candidate QM regions to $n^\mathrm{region}_\mathrm{max} = 250$ atoms, and the target QM region size to $n^\mathrm{region}_\mathrm{target} = 300$ atoms. Furthermore, we choose the minimum value for the search radius $r_s$ as $4\,\si{au}$, the cutting probability $p_\mathrm{cut}$ as $0.9$, and the number of reference QM regions $N^\mathrm{ref}$ as $4$. The radius $r_s$ is incremented in steps of $0.1\,\si{au}$, and $50$ candidate QM regions are generated for each value of $r_s$. Often large QM regions are required to converge relative energies in QM/MM.\cite{Rossbach2017,Liao2013,Sumowski2009} Since we will focus on the correlation treatment in the QM region rather than predicting observables, such as binding free energies, we selected the key parameters $n^\mathrm{region}_\mathrm{max}$ and $n^\mathrm{ref}_\mathrm{max}$ as a compromise between computational cost and accuracy. A dedicated uncertainty analysis will be required to propagate uncertainties from the QM region selection to the target properties (e.g., binding free energies).

To avoid unphysical fragments in the QM calculations, we cap the QM region with hydrogen atoms at all bonds where a covalent bond was cut between the QM and MM regions. We account for the polarization of the QM region by the MM region through electrostatic embedding\cite{Warshel_second-ONIOM-1976} of the QM region by the MM point charges. To avoid charge leaking from the QM region into the MM region caused by closely positioned MM charges, the charges are redistributed at the hydrogen caps based on the $RC$ scheme in Reference~\citenum{Lin2005}. We use the program SWOOSE\cite{Brunken2020, Brunken2024, Brunken2021} to perform the capping of the QM region and redistribute the MM point charges. We take the MM point charges from the Amber99sb*ILDN protein force field\cite{Lindorff‐Larsen2010} and the TIP3P water model\cite{Jorgensen1983} employed in the original MM-MD simulation\cite{paper1}.

\label{subsec:biological-system}
\subsection{BE Calculations}
We perform all BE calculations using \texttt{QuEmb}\cite{oimeitei/quemb:-2024-09-03, softwarepaper}, a Python package based on \texttt{PySCF}\cite{pyscf}. Github repository \texttt{mixed-basis-be}\cite{lweisburn/mixed-basis-be:-2024-11-19} includes both QM/QM/MM routines and mixed-basis BE functions such that the calculations introduced in this work can be replicated. For all calculations, we use CCSD as the high-level solver, though other wavefunction methods in principle can be used. 
\label{subsec:be_details}

\section{Results}\label{sec:results}
\subsection{Small 19G Molecule in MM Environment}\label{subsec:qm/mm}

Various snapshots of the 19G molecule in its aqueous environment are used to assess the accuracy of the QM/QM/MM method. These structures are not generated with automatic region selection, and no bonds are disrupted between the QM and MM regions. The MM region is treated as described in Section \ref{subsec:biological-system}. CCSD energies obtained for QM/MM Hamiltonians dressed with MM charges are the target references. We evaluate BE$(1)$, BE$(2)$, and BE$(3)$ energies for a subset of the said QM/MM Hamiltonians following the procedure outlined above.
We use the relatively small def2-SVP basis set for the CCSD and BE$(n)$ calculations so that we can compute reference values.

\begin{figure}[tb]
    \centering
    \includegraphics[width=0.9\linewidth]{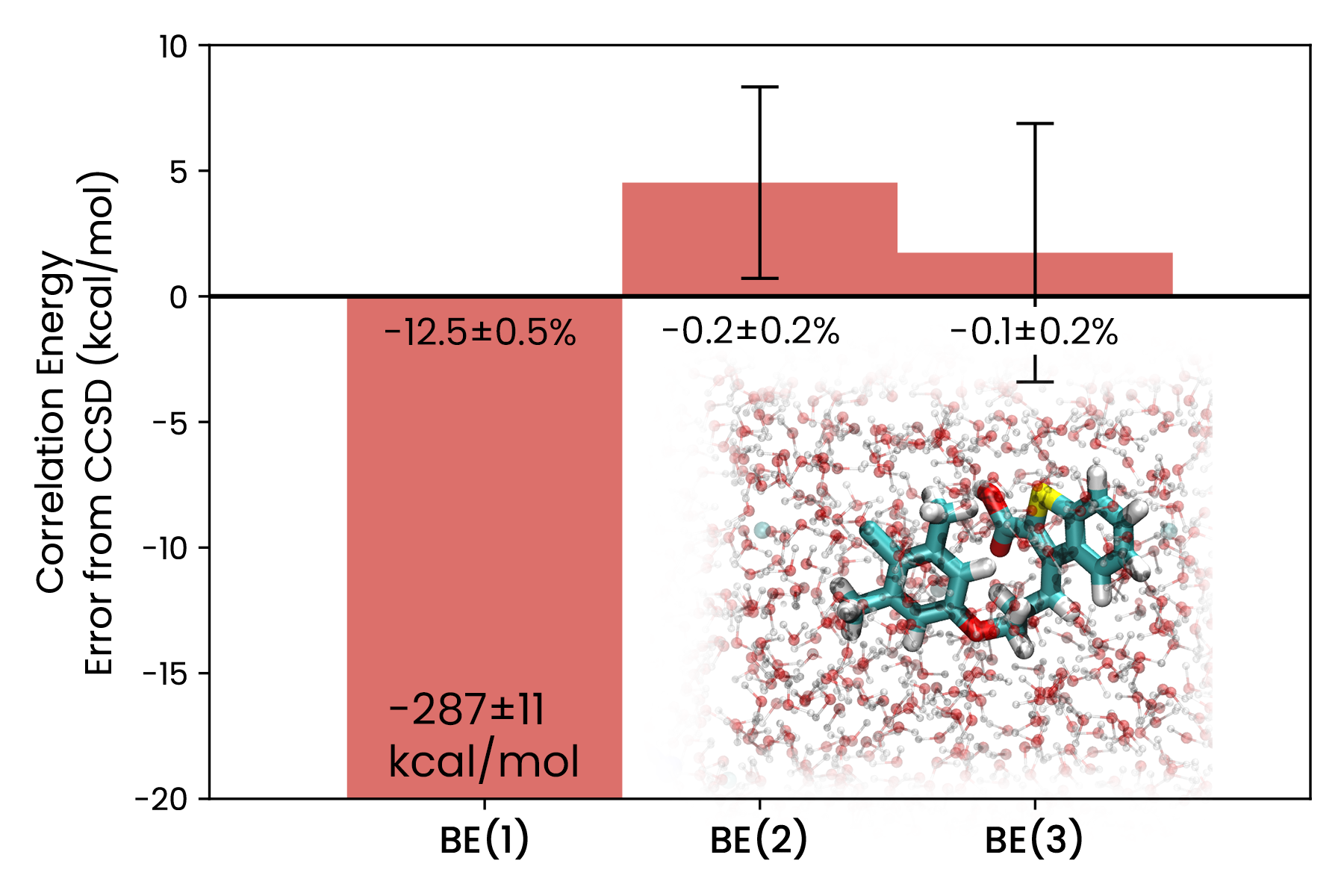}
    \caption{Correlation energy (CE) errors (kcal/mol) for BE$(1)$, BE$(2)$, and BE$(3)$ for 19G in water compared to QM/MM CCSD CEs for the QM region. The percent error in CE for each of these systems, averaged across ten structures, is given. One configuration in solvent is shown. This shows the expected convergence with increasing fragment size.}
    \label{fig:ligand}
\end{figure}

As shown in Figure \ref{fig:ligand}, BE with CCSD solver provides a robust approximation to CCSD results for these chemical Hamiltonians with electrostatic MM interactions. The percent errors in the correlation energy here are averaged over ten separate configurations across the MD trajectory. These accurate results are in line with previously reported data for systems without MM charges,\cite{iao-paper} where BE$(1)$ produces about 10\% error from CCSD, and BE$(2)$ and BE$(3)$, sub-0.5\%. We verify that BE is a valid strategy that one may use to further reduce the size of QM-MM Hamiltonians for resource-limited circumstances, such as applications on near-term quantum computers, while retaining high accuracy. This allows for expanded, chemically-motivated QM regions which are traditionally out of reach of wavefunction methods to be chosen.

\subsection{Mixed-Basis BE for Polyacetylene and Polyglycine Chains}

Following the procedure in Section \ref{sec:mixed-basis}, we evaluate energies of linear E-polyacetylene chains of varying lengths. Standard BE$(n)$, which uses a large basis for the entire system, serves as the target for the mixed-basis BE$(n)$-in-BE$(m)$ results. We first choose polyacetylene as a test conjugated hydrocarbon system for which we may easily obtain both conventional BE and CCSD references for a range of system sizes and basis sets.

\begin{figure*}[tb]
    \centering
    \includegraphics[width=0.9\linewidth]{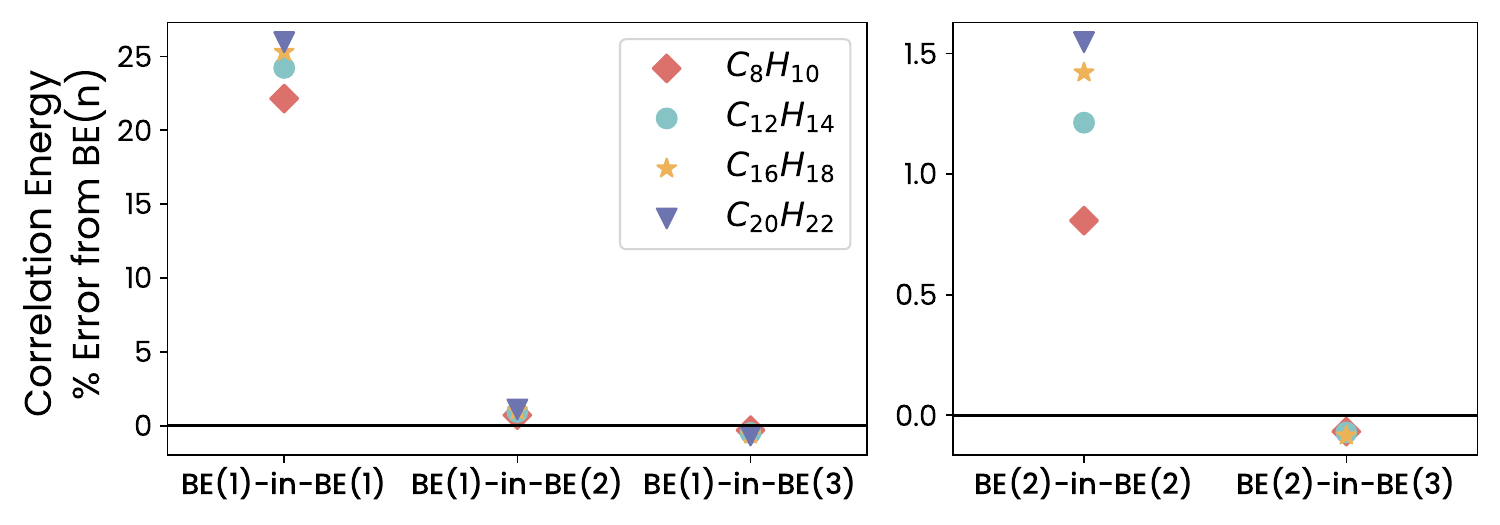}
    \caption{Percent errors of BE$(1)$-in-BE$(m)$ (left) and BE$(2)$-in-BE$(m)$ (right) CEs relative to the BE$(1)$ or BE$(2)$ CEs for E-polyacetylene chains of various lengths in the cc-pVDZ basis set.}
    \label{fig:polyac_dz}
\end{figure*}

Figure \ref{fig:polyac_dz} shows the percent error in the correlation energy compared to the standard BE results for molecules of increasing length with the cc-pVDZ basis set. That is, for BE$(n)$-in-BE$(m)$, we evaluate the percent error as follows, where CE is the correlation energy.

 \begin{equation}
     \text{Percent Error} = 
     100 \times\frac{
     \text{CE}_\text{BE$(n)$-in-BE$(m)$} - \text{CE}_{\text{BE$(n)$}}}
     {\text{CE}_{\text{BE$(n)$}}}
 \end{equation}
 
\begin{figure*}[tb]
    \centering
    \includegraphics[width=0.9\linewidth]{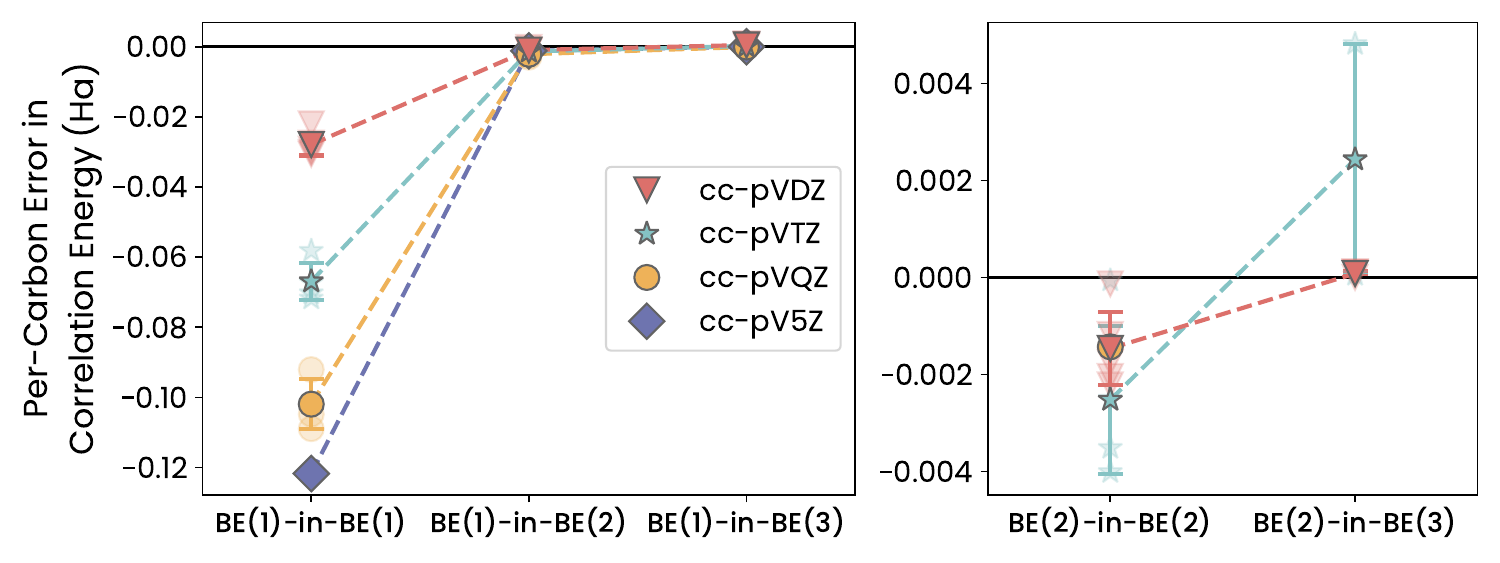}
    \caption{Geometry-averaged per-carbon error for BE$(1)$-in-BE$(m)$ schemes compared to BE$(1)$ (left) and BE$(2)$-in-BE$(m)$ convergence to BE$(2)$ (right) for E-polyacetylene molecules.}
    \label{fig:epoly-all-basis}
\end{figure*}

We observe that there is a robust convergence of BE$(1)$-in-BE$(m)$ with increasing $m$ (large basis set region size) to the standard BE$(1)$ result, despite the reduced computational cost from using a small fraction of the basis functions. BE$(2)$-in-BE$(2)$ already closely approximates the BE$(2)$ correlation energies, reaching about 1.5\% error from BE$(2)$. For the tested systems, mixed-basis calculations provide a robust approximation to the sub-1\% errors of large basis BE$(2)$ calculations relative to CCSD references. We provide geometry-averaged, per-carbon results in Figure \ref{fig:epoly-all-basis} to show that the convergence behavior remains comparable for the different choices of the basis set. By showing per-carbon error, we can compare systems of different lengths across these basis sets, rather than showing a single calculation of fixed length at each, to evaluate the method behavior. We note that BE$(n)$-in-BE$(n)$ has a notable degradation with larger basis (shown in Figure \ref{fig:epoly-all-basis}) and longer chain length(Figure \ref{fig:polyac_dz}), which is discussed further in the Supporting Information alongside the full results data. We discuss in Section \ref{subsec:ect} this behavior specifically in the case of BE$(1)$-in-BE$(1)$ and propose future work to reduce the deviation.

We choose polyglycine as an additional test because of its heterogeneous constituent element makeup. Unlike polyacetylene chains, polyglycine molecules contain the nitrogen and oxygen commonly found in biomolecules and introduce the challenge of non-uniform fragments. While obtaining CCSD references for longer polyglycine chains in larger basis sets is less trivial, we can verify the efficacy of the mixed-basis BE method on molecules more directly relevant to biological applications by probing polyglycine chains of limited length. 

\begin{figure*}[tb]
    \centering
    \includegraphics[width=0.9\linewidth]{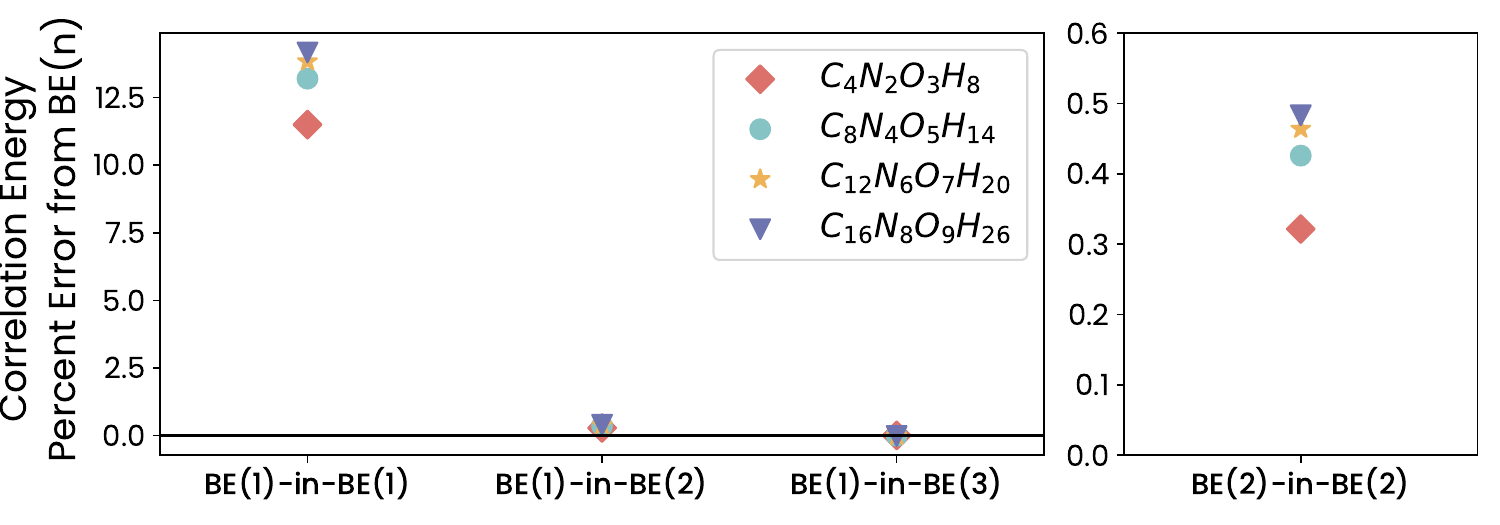}
    \caption{Percent errors of BE$(1)$-in-BE$(m)$ (left) and BE$(2)$-in-BE$(2)$ (right) CEs relative to the BE$(1)$ or BE$(2)$ CEs for polyglycine chains of various lengths in the cc-pVDZ basis set.}
    \label{fig:polygly}
\end{figure*}

As shown in Figure \ref{fig:polygly}, we observe the convergence behavior of BE$(1)$-in-BE$(m)$ to BE$(1)$ witnessed with the polyacetylene chain in polyglycine chains as well. As before, we confirm that BE$(2)$-in-BE$(2)$ produces a sub-1\% error relative to the BE$(2)$ correlation energies, which is close to the CCSD results. Positive results with the polyglycine chain encourage us to apply the schemes to realistic biological systems we introduce below to generate large basis set CCSD-level results.

\subsection{Mixed-Basis BE for Large Biological Systems}

We apply the combined QM/QM/MM and mixed-basis BE approach to a biological system to demonstrate its efficacy. We obtain the structures from an MD trajectory of the MCL1-19G complex in an explicit water solvent. The automatic QM region selection routine described in Section \ref{subsec:qm-selection} facilitates the construction of the QM Hamiltonians that account for the MM charges. Among approximately 17000 atoms in the solvated protein-ligand system, we choose about 350 atoms to treat on a quantum mechanical level. The system generation is described in Section \ref{subsec:biological-system}.

To model application on near-term quantum devices, which require a small fragment size to be solved with a limited number of qubits, we perform BE$(1)$-in-BE$(2)$ calculations to obtain the correlation energies. We use def2-TZVP/STO-3G as the high-level/minimal basis sets for the system. We note that treating the entire QM region using a large basis (i.e. def2-TZVP) and a correlated solver is beyond reasonable computing resources. Therefore, CCSD-level reference results are not available for this set of structures. This biological system, however, offers an opportunity to assess how the information the QM/QM/MM calculations captures beyond cheaper mean-field approaches. We ask if we can use the HF energies to predict the correlation energies obtained from BE$(n)$ calculations, in which case relatively costly correlated calculations have less practical value. Only if the correlation between HF energies and correlation energies are weak, we may justify a correlated treatment of biological systems using classical or quantum algorithms.

\begin{figure*}[tb]
    \centering
    \includegraphics[width=0.85\linewidth]{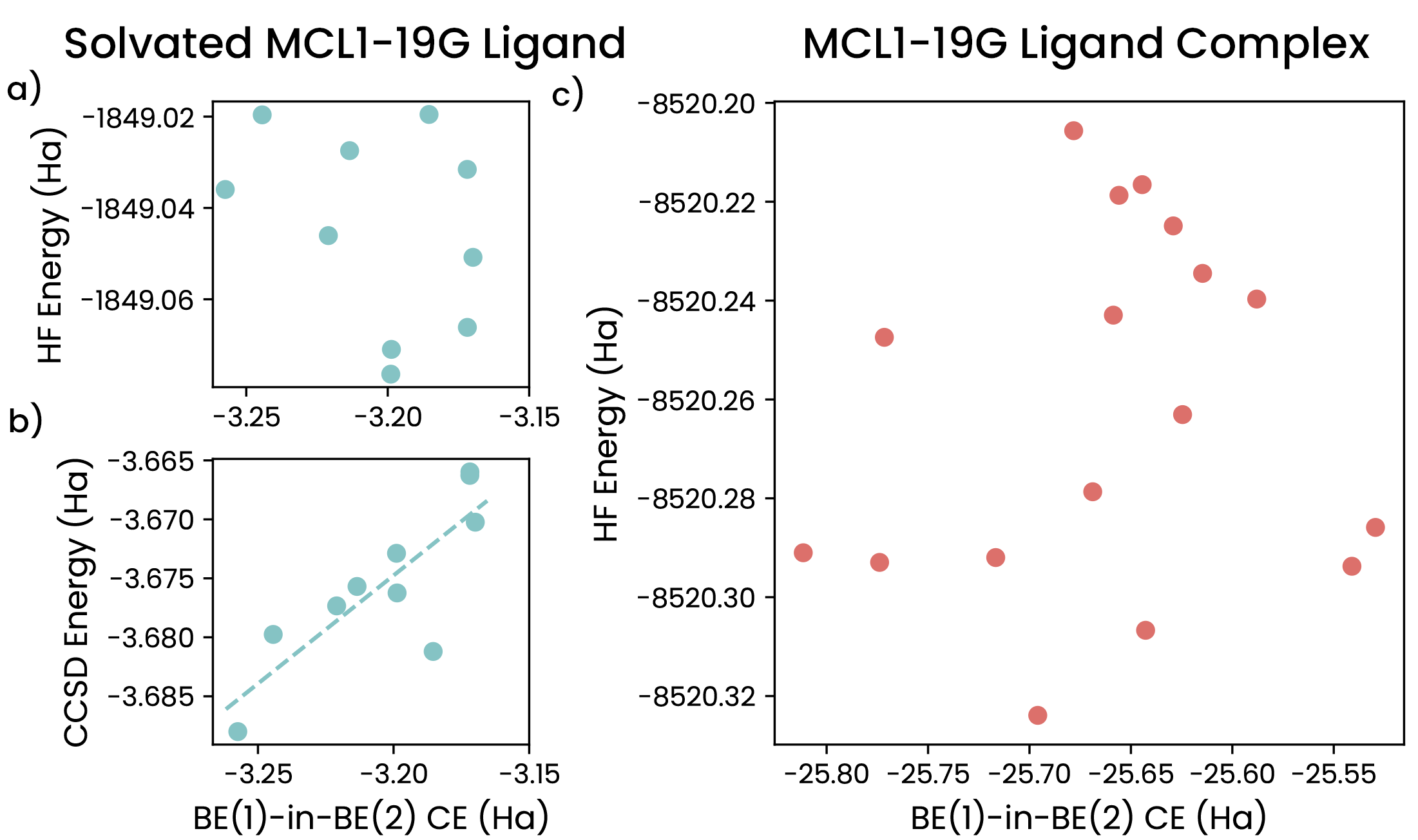}
    \caption{BE$(1)$-in-BE$(2)$ CEs for the solvated 19G ligand in the def2-SVP basis (left) and for MCL1-19G complex with the def2-TZVP basis (right). The structures included here are sampled from MD trajectories for both systems. Energies are compared to a) HF energy, b) CCSD energy, and c) HF energy for the QM/MM system.}
    \label{fig:complexcorr}
\end{figure*}

The low degree of correlation exhibited in the right panel of Figure \ref{fig:complexcorr} suggests that one cannot determine the BE$(1)$-in-BE$(2)$ correlation energies from the HF energies. To assess the accuracy of BE$(1)$-in-BE$(2)$ in this context, we look to the solvated 19G ligand MD structures from Section \ref{subsec:qm/mm}, in the def2-SVP basis, for which we can obtain CCSD references. As shown in the left panel of Figure \ref{fig:complexcorr}, BE$(1)$-in-BE$(2)$ is a good proxy of CCSD for this set of much reduced structures, with energies also uncorrelated to HF results. As we will further discuss in Section \ref{subsec:ect} with Figure \ref{fig:ect-corre}, BE$(1)$-in-BE$(2)$, unlike BE$(1)$-in-BE$(1)$, exhibits slow degradation in the correlation energy error with respect to the error in electron count. For large complexes with dense basis, where electron count error becomes pronounced, we predict that BE$(1)$-in-BE$(2)$ remains as a cost-effective correlated model where one can access electron correlation effects beyond HF in large systems using attainable amount of computational resources.

While we present here the single-point correlation energies of an ensemble of structures, the energies obtained with this QM/QM/MM approach can be further used alongside non-equilibrium switching\cite{Kearns2017} and other techniques to obtain the system binding free energies. The free binding energy for this system requires extensive calculations beyond the scope of this work, but such a method has been described at length in other studies.\cite{paper1,paper3a} To this end, this correlation and its more accurate calculation are crucial to understand drug binding, where interactions beyond the mean-field can be important. Per our benchmark studies, the estimation of these energies improves with increasing fragment size as quantum hardware improves, leading to a more accurate depiction of systems of interest with increased computational resources.

\section{Discussion}\label{sec:discussions}
\subsection{Mixed-Basis BE Scaling}\label{subsec:mb_be_scaling}

\begin{figure*}[tb]
    \centering
    \includegraphics[width=0.9\linewidth]{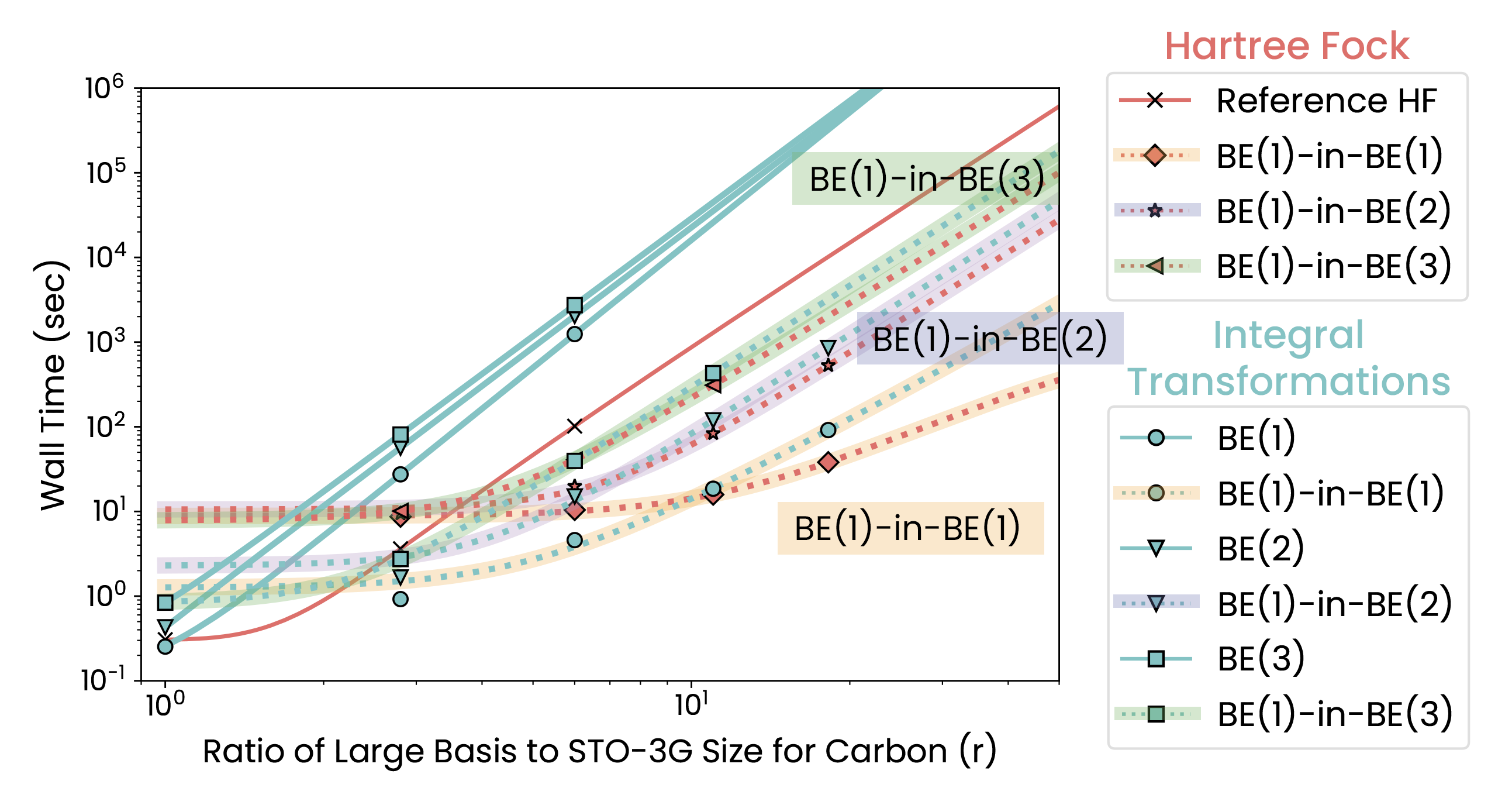}
    \caption{Empirical scaling for $C_{16}H_{18}$ HF calculations and integral rotations. This is the total time for all reference HF calculations and for all integral transformations into all fragment spaces for the given system. BE$(n)$ timings are fit with a solid curve, while BE$(n)$-in-BE$(m)$ are fit with dotted curves. BE$(1)$-in-BE$(1)$ is highlighted in orange, BE$(1)$-in-BE$(2)$ is highlighted purple, and BE$(1)$-in-BE$(3)$ is highlighted green. For integral transformation timings, BE$(m)$ and BE$(n)$-in-BE$(m)$ data share a marker shape. All timings are fit to the form $t = A*r^4 + B*r^3 + C$, except for the BE$(n)$ integral transformation timings fit to $t = A*r^B +C$.} 
    \label{fig:empirical}
\end{figure*}

As discussed in Section \ref{sec:mixed-basis}, the BE$(n)$-in-BE$(m)$ method requires a separate mixed-basis HF calculation and fragment Hamiltonian for each of the BE$(m)$ regions. This differs from conventional BE calculations, which use a single large-basis HF reference to compute all fragment Hamiltonians. Here, we quantify the difference in the computational scaling between the conventional BE calculation and the proposed mixed-basis variant, identifying the regime in which the mixed-basis BE calculations prove more computationally efficient.

First, we analyze the efficiency of the methods as we change the density of the basis set (the number of basis functions per the atom, $N_\text{bf}$) while keeping the total number of atoms fixed. As we will see, the governing principles in this respect will be relatively insensitive to the details of the implementation. We will then discuss in general terms the scaling with the number of atoms, which will be more sensitive to the implementation details.

As a demonstrative example, we will focus on the timing results for a simple, homogeneous, 16 carbon E-polyacetylene chain. This allows us to obtain a complete set of reference BE$(n)$ timings with relatively large basis sets, which are compared with BE$(n)$-in-BE$(m)$ timings in various basis sets in Figure \ref{fig:empirical}. For both the HF calculation and the formation of the fragment Hamiltonians, the bottleneck is four-index integral transformations, whose computational cost formally scales as $\mathcal{O}(N_\text{bf}^4)$. Accordingly, if one moves from a STO-3G basis ($N_\text{bf} \approx 4$) to a cc-pVTZ basis ($N_\text{bf} \approx 30$) for standard large-basis BE$(n)$, one expects a calculation of both the HF energy and the fragment Hamiltonians to slow by a factor of $\left(\frac{30}{4}\right)^4 \approx 3000$, a substantial slowdown. Indeed, we observe a slowdown in both steps with changing basis set in the calculation as shown by the solid curves, representing BE$(n)$ data, in Figure \ref{fig:empirical}. Note that the integral transformation steps, in solid blue, are similarly expensive for BE$(1)$, BE$(2)$, and BE$(3)$ for a given basis set.

Meanwhile, in the mixed-basis calculations, only a handful of the atoms are in large basis. For large systems especially, each of the mixed-basis calculations take only slightly longer than the minimal basis calculations. Of course, we must repeat the mixed-basis calculations independently for each atom-centered BE$(m)$ fragment. As long as the number of atom-centered BE$(m)$ fragments is smaller than the speed-up achieved by mixed-basis formulation compared to large-basis calculations, we expect the BE$(n)$-in-BE$(m)$ calculations to be faster than BE$(n)$. This effect is indeed observed for the HF portion, with orange lines, of the calculations in Figure \ref{fig:empirical}: for the E-polyacetylene chain in the cc-pVTZ basis set, a complete set of BE$(n)$-in-BE$(m)$ HF calculations is approximately 10 times faster than the single large-basis HF calculation required for BE$(n)$.

For the fragment Hamiltonian evaluation, the mixed-basis BE$(n)$-in-BE$(m)$ achieves even better computational cost savings compared to large basis BE$(n)$. Much of the work required to compute the Hamiltonian for fragment $A$ is independent of the work required for fragment $B$. Because the Hamiltonian for fragment $A$ (and respectively, $B$) deals with orbitals near its center atom, integral transformation for fragment $A$ typically requires a set of local integrals that overlaps very little (or not at all) with the integrals of fragment $B$. As a result, the cost of evaluating the fragment Hamiltonians is largely independent to whether a single or multiple HF references are used. In mixed-basis BE$(n)$-in-BE$(m)$ calculations, the use of smaller number of basis functions compared to BE$(n)$ has an uncompromised speedup, unlike the HF step. This effect is observed in the blue line fragment Hamiltonian results in Figure \ref{fig:empirical}: the speedup factors for fragment Hamiltonian calculations are significantly larger than that for HF portions. For example, in a cc-pVTZ basis, we expect the complete set of BE$(n)$-in-BE$(m)$ fragment Hamiltonian evaluation to be approximately 3000 times faster in the limit where the BE$(m)$ fragments are small compared to the entire chemical system. In practice, for our example system in Figure \ref{fig:empirical}, we observe 30-300x speedup depending on the choice of $n$. Given that the generation of fragment Hamiltonians is generally much more time consuming than the HF calculations, we conclude that the mixed-basis BE$(n)$-in-BE$(m)$ will be faster than BE$(n)$ for essentially any basis set larger than minimal basis for most molecular sizes.

Our study of a small system (on the order of tens of atoms) and the analysis of larger ones (containing hundreds, or in the extreme, thousands of atoms) reveal three general regimes. In the first regime, for suitably \textit{small systems}, individual BE$(m)$ fragments will be a significant fraction of the full system. As a result, mixed-basis calculations will be similar in their computational cost to the large basis BE$(n)$ calculation and BE$(n)$-in-BE$(m)$ will not offer a significant speedup. We see this in the smallest basis set sizes shown, cc-pVDZ, for the 16 carbon system, where the set of BE$(n)$-in-BE$(m)$ HF calculations is more expensive than a single reference in cc-pVDZ. For \textit{moderate-sized systems} in the second regime, this smaller basis crossover does not exist, and BE$(n)$-in-BE$(m)$ will be much faster in all non-minimal basis sets. Our target biological systems like the MCL-19 complex likely belong to this regime. In the final regime, for \textit{very large systems}, performing $\mathcal{O}(N_\text{atom})$ separate HF calculations  become prohibitive, and BE$(n)$ will again be faster than BE$(n)$-in-BE$(m)$. The precise transitions between these regimes will heavily depend on the implementation of HF and fragment Hamiltonian routines.

In the present work, we rely on a conventional HF calculation and four index fragment integral transformations that loop over each fragment. As noted above, this means that the second regime, where mixed-basis BE prevails, extends to very large systems. Integral transformation, which is the main bottleneck in our present implementation, is faster with the mixed-basis formulation for all system sizes. Therefore, crossover to the third regime will only happen when the total time of fragment HF calculations in mixed-basis BE$(n)$-in-BE$(m)$ exceeds the time required to generate the fragment Hamiltonians in the large basis for BE$(n)$. Such situation will only occur for systems inaccessible to any conventional HF implementation. (e.g. $> 10,000$ atoms)

We note that there are, however, a number of ways in which the HF and fragment Hamiltonian calculations could be accelerated. This would in turn affect the system size at which the third regime begins. By exploiting the rank-deficient properties of the ERI tensors\cite{Strout_QuantStudyHFScaling-1995} with methods like resolution of the identity (RI) or density fitting (DF),\cite{Dunlap_PySCF-RI-2000, Ren_RI-2012, Feyereisen_ApproxIntegrals-1993, Jung_AuxBasis-2005} or more recent developments in tensor hypercontraction techniques,\cite{Hohenstein_THC-1-2012,Parrish_THC-2-2012,Hohenstein_THC-3-2012,Lu_ISDF-2015,Qin_ISDF-2023} the speedup between the mixed-basis and large-basis calculations can be reduced to $\mathcal{O}(N_\text{bf}^3)$. Furthermore, by exploiting sparsity\cite{L_Kussmann_2013}, the cost of large-basis HF and fragment Hamiltonian calculations can be reduced to $\mathcal{O}(N_\text{atom}^2)$ for large enough systems. Finally, the fragment Hamiltonian calculations can be accelerated by avoiding redundancy in overlapping fragments\cite{be-ccsd} at the expense of some loss in parallelism. Each of these improvements can lead to a substantial reduction in cost for both large-basis HF and fragment Hamiltonian evaluation for BE$(n)$. These changes, however, will have a less pronounced impact on the cost of mixed-basis BE$(n)$-in-BE$(m)$ calculations in the limit of large systems. Thus, we expect that a more sophisticated implementation will bring the third regime closer to an achievable system size. However, without implementing these algorithms, it is difficult to predict where the crossover will occur. Implementation of these improvements will be a subject of future work. For the time being, we simply note that the simple expedient of using mixed basis sets in BE$(n)$-in-BE$(m)$ significantly extends the system sizes one can access with conventional techniques while requiring little code modification.

\subsection{Electron Count and Self-consistency}
\label{subsec:ect}

\begin{figure}[tb]
    \centering
    \includegraphics[width=0.9\linewidth]{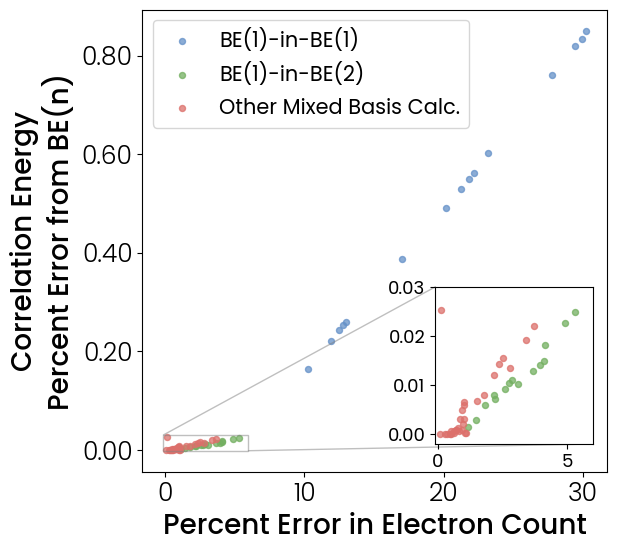}
    \caption{Correlation plot of the percent errors in electron count and in CEs (relative to BE$(n)$) from mixed-basis BE calculations on polyacetylene chains. We see a strong linear relationship between electron count error and percent CE error for BE$(1)$-in-BE$(1)$ relative to BE$(1)$ and a much weaker relationship for other mixed-basis set calculations.}
    \label{fig:ect-corre}
\end{figure}

In BE$(1)$-in-BE$(1)$, we observe a large systematic underestimation of the correlation energies compared to BE$(1)$. In these calculations, the center-site orbitals that factor into the final energy expression are surrounded only by orbitals consisting of minimal basis set functions. The large flexible basis functions used to build the fragment orbitals shifts the energy of the fragment such that a net flow of electrons onto the fragment is favored. As shown in Figure \ref{fig:ect-corre}, there is a strong correlation between the percent-error of the electron count (evaluated as in Equation \ref{eq:ect}) and the percent-error of mixed-basis BE$(n)$-in-BE$(m)$, as compared to its full BE$(n)$ counterpart. 

    \begin{equation}
        \label{eq:ect}
        N_{\text{electron}} = \sum_F \sum_{i\in\mathbb{C}^F} P_{ii}^{F}
    \end{equation}

Here, $P^{F}$ is the 1-RDM from the correlated calculation of fragment $F$, and $i\in\mathbb{C}^F$ are the center-site indices of fragment $F$.

As introduced in previous works\cite{be-mol, be-atom, be-ccsd} related to BE, adjustment of a global chemical potential $\mu$ can correct the uneven energetic stabilization of the flexible fragment orbitals. In practice, this is represented as a single parameter $\mu$ that is optimized by iteratively performing BE calculations with a dressed Hamiltonian described in Equation \ref{eq:pot_ham}
\begin{equation}\label{eq:pot_ham}
    \Hat{H}_\text{emb}^F \leftarrow \Hat{H}_\text{emb}^F + \mu\sum_{i\in\mathbb{C}^F} a_i^{F\dagger}a_i^F
\end{equation}
until the electron count evaluated with Equation \ref{eq:ect} matches the number of electrons physically present in the chemical system of interest.

We expect that BE$(n)$-in-BE$(m)$, and especially BE$(1)$-in-BE$(1)$, will benefit from such a self-consistency loop. This will be explored in future studies, especially for classical computing applications in which iterative BE can be more viable.

\section{Conclusions}\label{sec:conclusion}
We provide a novel framework of multiscale embedding to apply near-term quantum computers for realistic quantum chemical and biological questions of interest. We successfully show for the first time that we can couple traditional QM/MM embedding and BE to form small, tunable regions amenable for quantum computers or classical correlated solvers like CCSD with limited computational resources. 
The convergence of BE correlation energy to the full-system reference with fragment size naturally lends itself to both NISQ devices, which have fewer qubits and can be used to solve small fragments, and future technology with an increasing number of usable qubits.

We further develop and demonstrate a new mixed-basis BE method, which allows for these calculations to be performed with great cost reduction and controlled loss of accuracy. We finally show that, coupled with automatic QM region selection techniques, this layered method can be applied to biological systems of interest with tens of thousands of atoms, and hundreds of atoms in the QM region, to generate CCSD-level results. This offers a new method to explore electron correlation effects in these important fields.

In future work, we aim to investigate several of the questions introduced above. The mixed-basis embedding framework introduces error to the density matrix reassembled from the fragments; a study of this error and its correction can further improve the method's accuracy and application. Applying a self-consistency loop between the classical optimizer and fragment calculations that enforces the correct number of electrons using global chemical potential and density matching could improve the accuracy of the correlation energies from mixed-basis embedding calculations on either classical or quantum computers. Faster HF reference calculations and improved integral transformations with density fitting and other techniques in addition to this framework can further lead to the study of even larger systems. This work also leads to application and simulation directly on quantum computers, which is investigated in other works,\cite{paper2, paper4, paper5} as well as additional thorough study of biological systems.\cite{paper1,umbrella} To this purpose, for classical simulation with larger overlapping fragments, we look to examine other non-connectivity based fragmentation schemes to address explicit solvents and non-bonded interactions.

While this work focuses on the embedding of chemical structures to calculate single point energies, the method can be used alongside other techniques to apply quantum computers to model realistic biochemical systems. This QM/QM treatment can be applied atop careful QM/MM studies of biological systems to better achieve more accurate drug binding energies.\cite{paper1} Other work shows that the BE method produces Hamiltonians whose ground states can be readily prepared on quantum computers.\cite{paper2} This paper, alongside future effort in classical biological simulation and quantum computing, provides a promising route forward to use emerging quantum computing technologies and algorithms on classically challenging problems.

\begin{acknowledgements}
This work is part of the research project ``Molecular Recognition from Quantum Computing'' and is supported by Wellcome Leap as part of the Quantum for Bio (Q4Bio) program.

Authors thank Dr. Victor Bastidas and Dr. Aram Harrow for many helpful conversations about the latest developments in quantum computing algorithms and hardware. Authors also thank Dr. Thomas Weymuth for insightful discussions about the biological systems as well as integration of the multiscale embedding techniques to high-throughput pipeline.
\end{acknowledgements}

\bibliography{main}
\bibliographystyle{apsrev4-1}

\end{document}


\maketitle

\section{Introduction}

We present the wall times for the Hartree Fock and integral transformation steps in Section \ref{Scaling}. 
We include more details on the QM region selection in Section \ref{qm-reg-selec}.
We present and explain the cumulant-based BE energy expression in Section \ref{energy-expr}.
We conclude with additional mixed-basis embedding results in multiple basis sets for E-polyacetylene and polyglycine molecules in Section \ref{Results}.

\section{Computational Time Scaling}\label{Scaling}
\begin{figure}[hp!]
    \centering
    \includegraphics[width=0.8\linewidth]{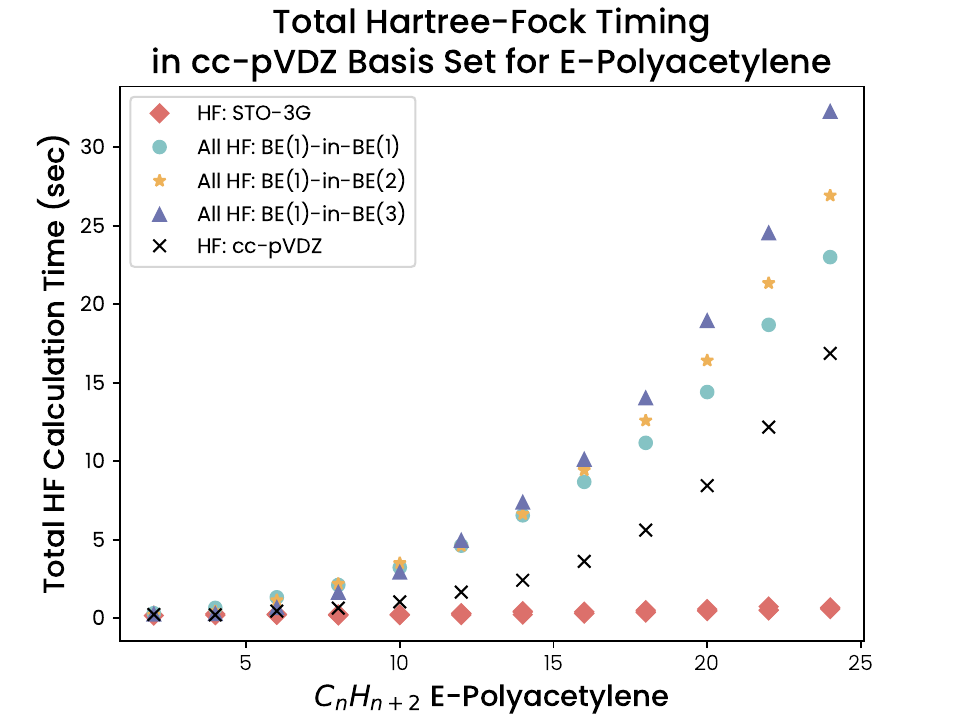}
    \caption{Computational wall times to complete the mean-field step for mixed-basis vs conventional BE for the cc-pVDZ basis. BE$(n)$-in-BE$(m)$ signifies the total time for all the reference mixed-basis cc-pVDZ/STO-3G HF calculations performed for a given $n$ carbon E-polyacetylene molecule. In this regime, while the memory requirement is lessened for the mixed-basis BE, the increased total number of reference calculations makes this step take longer to complete.}
    \label{fig:ccpvdz_hf}
\end{figure}

\begin{figure}[hp!]
    \centering
    \includegraphics[width=0.8\linewidth]{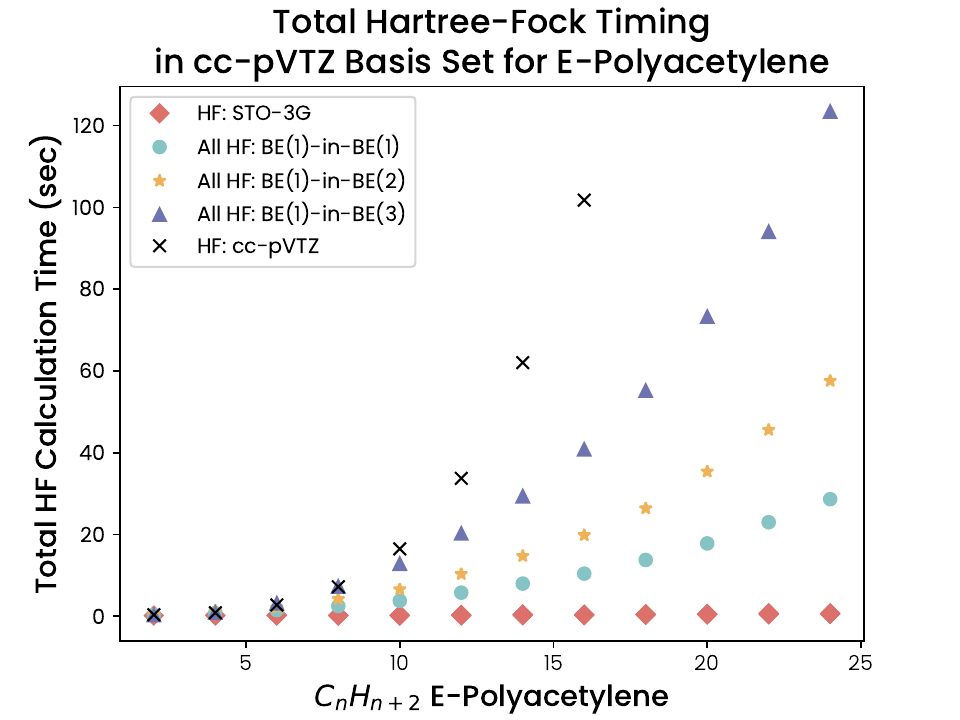}
    \caption{Computational wall times to complete the mean-field step for mixed-basis vs conventional BE for the cc-pVTZ basis. BE$(n)$-in-BE$(m)$ signifies the total time for all the reference mixed-basis cc-pVTZ/STO-3G HF calculations performed for a given $n$ carbon E-polyacetylene molecule. This shows the trade-off in timing for this step with increasingly large systems: while the time for the reference HF calculations to performed is similar between methods in the small molecule regime, mixed-basis BE requires less time to complete than standard BE with increasing E-polyacetylene length.}
    \label{fig:ccpvtz_hf}
\end{figure}

\begin{figure}[hp!]
    \centering
    \includegraphics[width=0.8\linewidth]{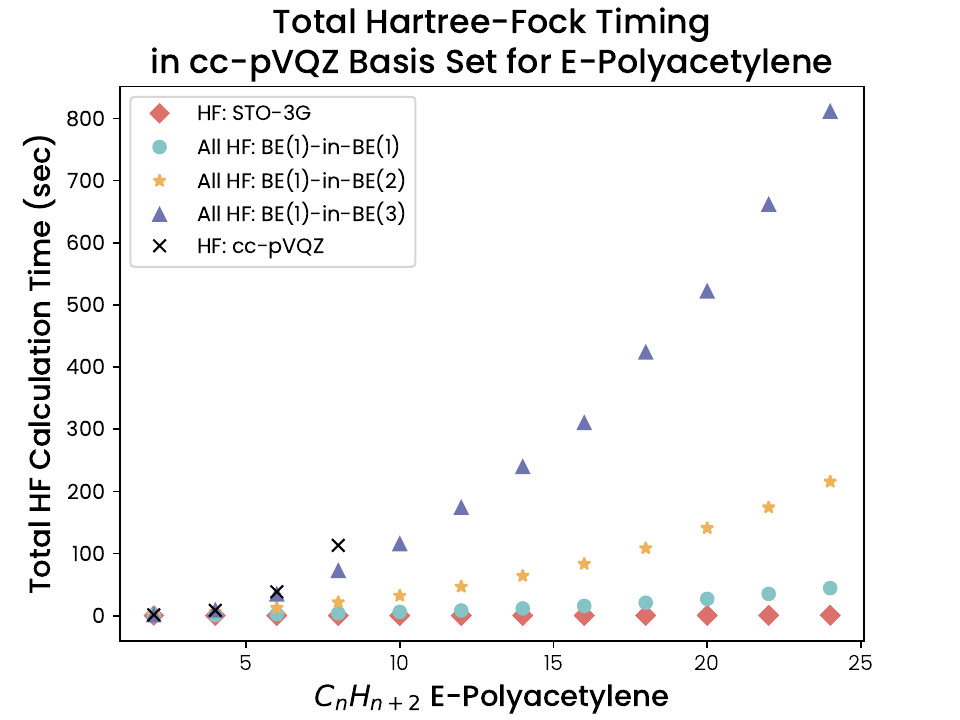}
    \caption{Computational wall times to complete the mean-field step for mixed-basis vs conventional BE for the cc-pVQZ basis. BE$(n)$-in-BE$(m)$ signifies the total time for all the reference mixed-basis cc-pVQZ/STO-3G HF calculations performed for a given $n$ carbon E-polyacetylene molecule. The large-basis BE reference is quickly infeasible due to large memory requirements, while mixed-basis BE can be performed for larger systems.}
    \label{fig:ccpvqz_hf}
\end{figure}

\begin{figure}[hp!]
    \centering
    \includegraphics[width=0.8\linewidth]{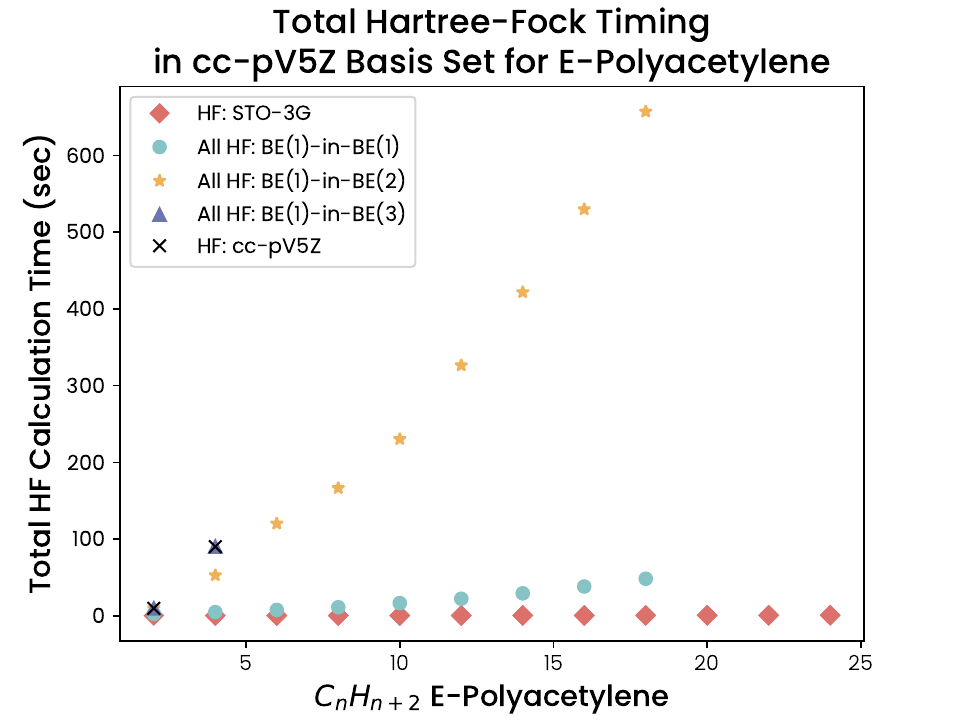}
    \caption{Computational wall times to complete the mean-field step for mixed-basis vs conventional BE for the cc-pV5Z basis. BE$(n)$-in-BE$(m)$ signifies the total time for all the reference mixed-basis cc-pV5Z/STO-3G HF calculations performed for a given $n$ carbon E-polyacetylene molecule. In this regime, while BE$(1)$-in-BE$(2)$ and BE$(1)$-in-BE$(3)$ are clearly significantly less costly than the large basis reference, the BE$(1)$-in-BE$(3)$ large-basis region is large enough that the calculation requires much more memory. For the 4 carbon E-polyacetylene, this large basis set region contains the whole molecule.}
    \label{fig:ccpv5z_hf}
\end{figure}

\begin{figure}[hp!]
    \centering
        \includegraphics[width=0.8\linewidth]{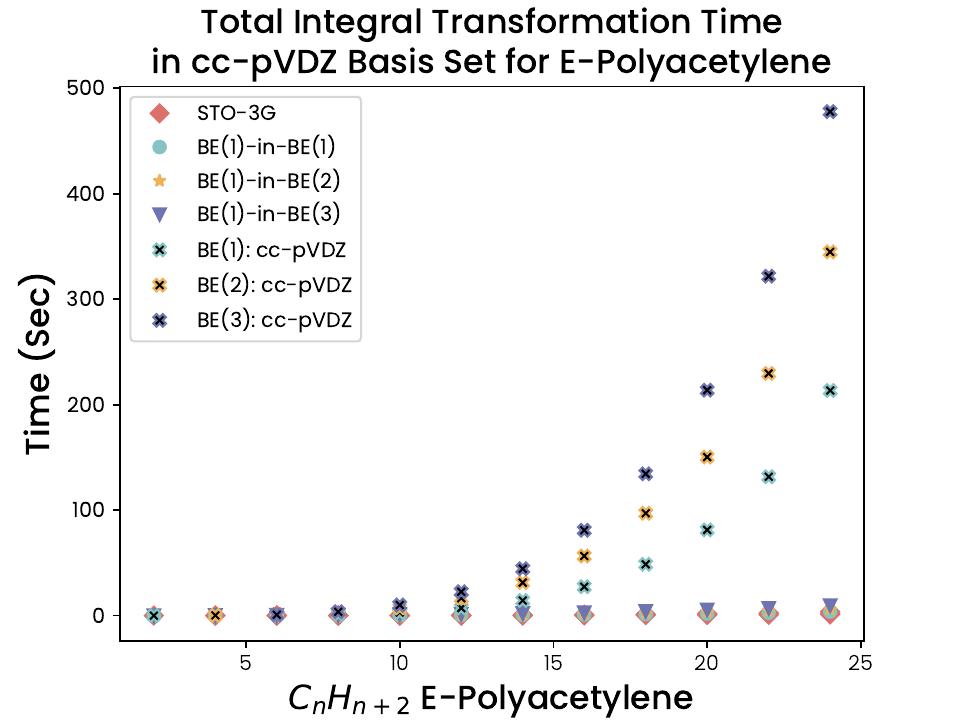}
    \caption{Computational wall times to complete the necessary two-electron integral transformations for mixed-basis vs conventional BE for the cc-pVDZ basis. In all cases, the timing for the integral transformation step in the mixed basis is less than for the conventional BE.}
    \label{fig:ccpvdz_ao2mo}
\end{figure}

\begin{figure}[hp!]
    \centering
        \includegraphics[width=0.8\linewidth]{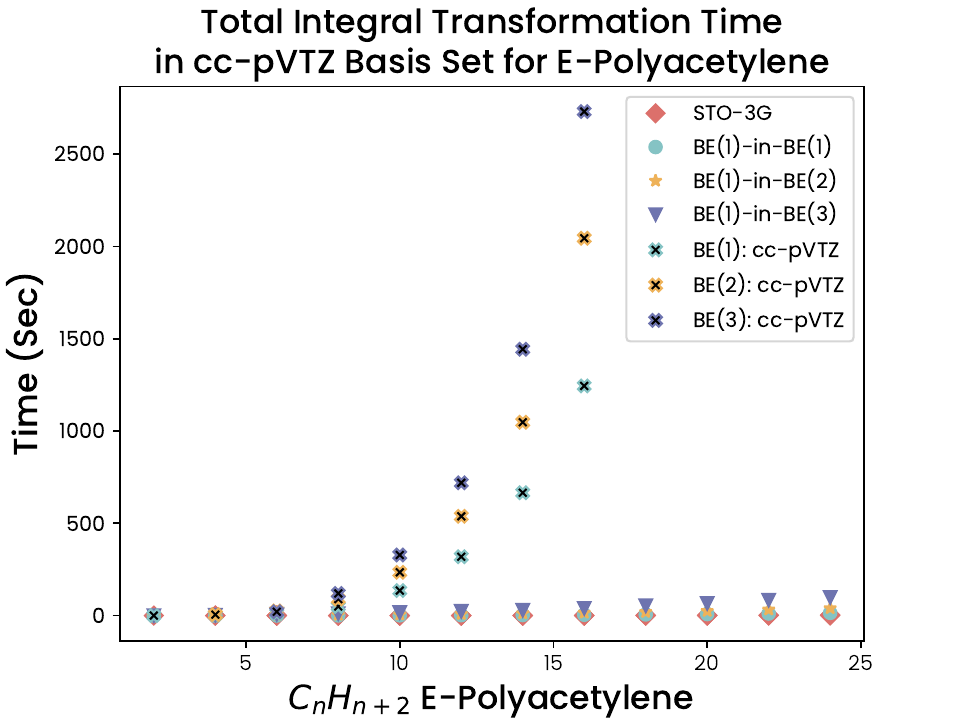}
    \caption{Computational wall times to complete the necessary two-electron integral transformations for mixed-basis vs conventional BE for the cc-pVTZ basis. Again, the timing for the integral transformation step in the mixed basis is less than for the conventional BE.}
    \label{fig:ccpvtz_ao2mo}
\end{figure}

\begin{figure}[hp!]
    \centering
        \includegraphics[width=0.8\linewidth]{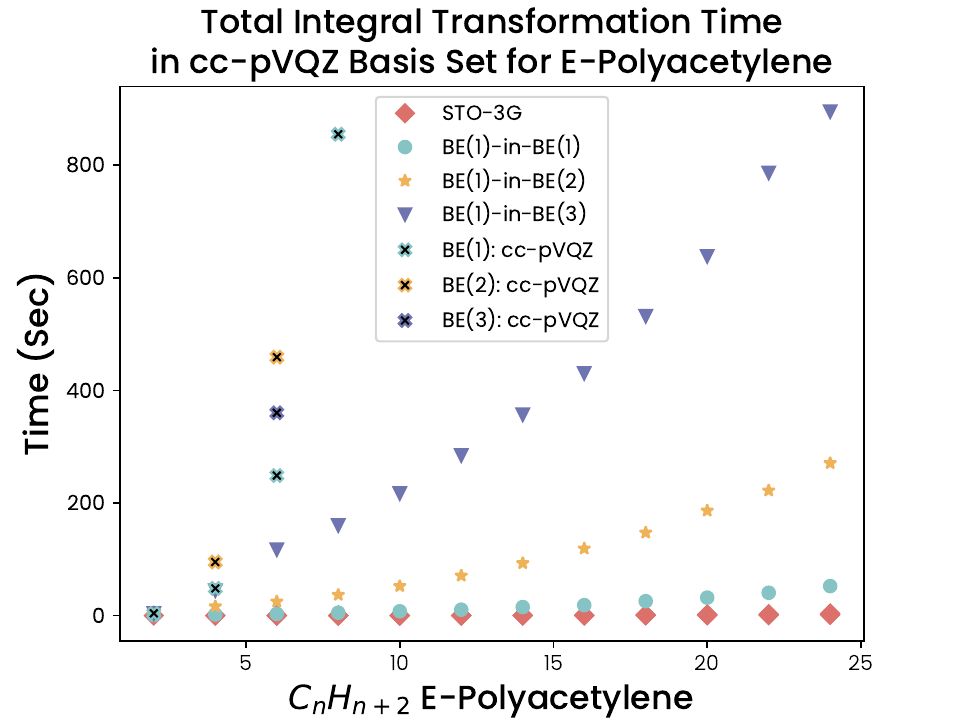}
    \caption{Computational wall times to complete the necessary two-electron integral transformations for mixed-basis vs conventional BE for the cc-pVQZ basis. Again, the timing for the integral transformation step in the mixed basis is less than for the conventional BE. In this case, traditional BE becomes quickly unviable.}
    \label{fig:ccpvqz_ao2mo}
\end{figure}

\begin{figure}[hp!]
    \centering
        \includegraphics[width=0.8\linewidth]{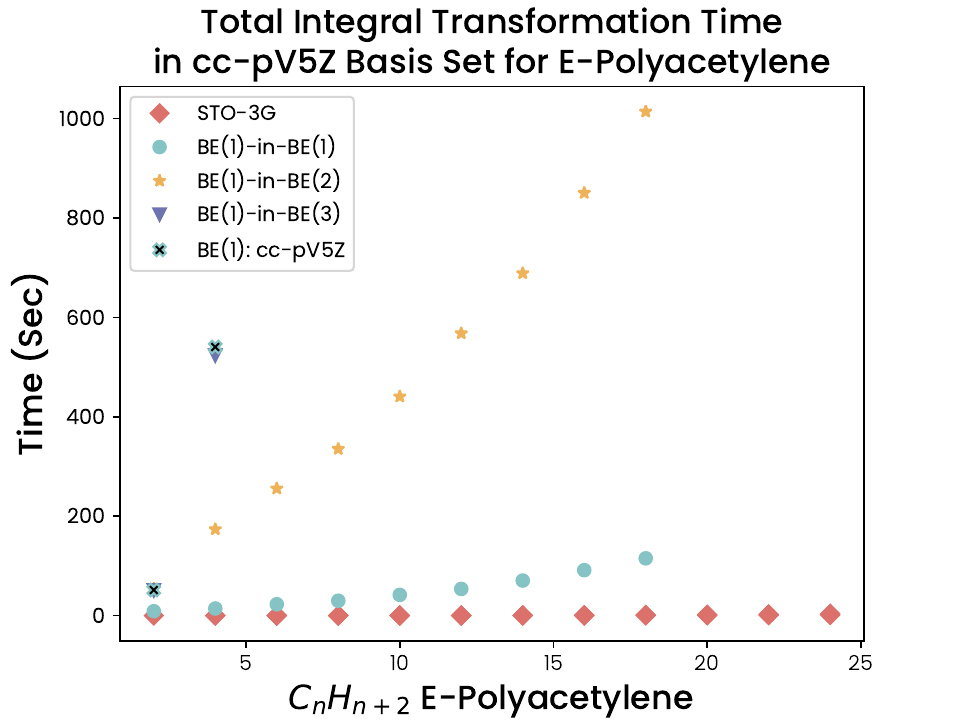}
    \caption{Computational wall times to complete the necessary two-electron integral transformations for mixed-basis vs conventional BE for the cc-pV5Z basis. Traditional BE is quickly impractical, though BE$(1)$-in-BE$(1)$ performs similarly to minimal STO-3G for this large basis set.}
    \label{fig:ccpv5z_ao2mo}
\end{figure}

Figures \ref{fig:ccpvdz_hf} - \ref{fig:ccpv5z_ao2mo} provide the wall times for steps in these mixed basis calculations to be completed, comparing the BE$(n)$ and the BE$(n)$-in-BE$(m)$ methods. For the standard BE$(n)$ calculations, we record the time for a single Hartree-Fock calculation to be performed. For the mixed-basis BE$(n)$-in-BE$(m)$ calculations, we report the total time for all mixed-basis HF reference calculations to be reported. For both methodologies, we report the total time for all integral transformations of the ERI's from the full atomic orbital AO space to the Schmidt space, one per fragment. These calculations were performed using \texttt{PySCF} with our in-house embedding code, \texttt{QuEmb}. They were performed on one node with 48 high memory 8 Gb cores on Intel Xeon Gold 5220R processors. All HF references were performed in serial for clarity in these timings, though this can be easily parallelized with some resource estimation.

\section{QM Region Selection}\label{qm-reg-selec}
For the large biological systems studied in this paper, we automate the QM/MM region selection using a novel algorithm described here.

For a given periodic structure $K$, we first translate all atoms in the structure by the periodic box vectors such that the ligand is in the center of all atoms. In practice, we shift the first ligand atom to the box center and translate all other atoms according to the periodic box vectors. Subsequently, we systematically construct a large number of QM region candidates by including all atoms within a given distance $r_s$ around the ligand atoms. We iteratively add MM atoms to these QM regions, which are connected by a covalent bond to a QM atom. In this way, we follow the covalent bonds until we arrive at a C--X bond with sp$^3$ hybridized C and $\text{X} = \text{C}, \text{N}$. These bonds are cut with the probability $p_\mathrm{cut}$ to separate the QM and MM region. We apply this probabilistic approach to generate an ensemble of different QM regions. We construct QM region candidates for different values of $r_s$ until all generated QM regions contain more atoms than $n^\mathrm{ref}_\mathrm{max}$, where $n^\mathrm{ref}_\mathrm{max}$ is the maximum number of atoms that we, with current computational resources, can use for the reference QM regions. From these QM region candidates, we take the largest $N^\mathrm{ref}$ QM regions as reference regions. For all other QM regions $Q^K_i$ of size $|Q^K_i| \leq n^\mathrm{region}_\mathrm{max}$, we calculate an error metric $\epsilon^\mathrm{force}_i$ from the QM/MM forces $f_{\alpha k}^i$ with $\alpha=x,y,z$ acting on the $N_\mathrm{ligand}$ ligand atoms and the average $\Bar{f}_{\alpha k}^\mathrm{ref}$ of the same forces from the QM reference regions,
\begin{align}
    \epsilon^\mathrm{force}_i = \frac{1}{3N_\mathrm{ligand}} \sum_{k=1}^{N_\mathrm{ligand}} \sum_{\alpha=x,y,z} \left| \Bar{f}_{\alpha k}^\mathrm{ref} - f_{\alpha k}^i \right|~.
\end{align}
We choose the QM region $Q_K = Q^K_i$ with the lowest $\epsilon^\mathrm{force}_i$ as the resulting QM region. We note that solvent atoms and solvated ions are excluded from the QM region in our approach because they are likely to diffuse through the simulation box, leading to a scattered QM region.

The QM region $Q^K_i$ is tailored to a single structure $K$, which may not be representative of an entire MD trajectory. Therefore, we perform this QM region selection process for a set of $N^\mathrm{selection}$ structures, which are equally spaced in the trajectory. The final QM region is constructed to include the $n^\mathrm{region}_\mathrm{target}$ most frequently selected atoms during the structure selection procedures for $K$. 

We achieve this final selection with the following algorithm:

\begin{enumerate}[label=(\roman*)]
\item We rank the atoms with index $k$ by the frequency $a_k$ in which they occur in the QM regions,
    \begin{align}
        a^\mathrm{atom}_k = \sum_K 1~\text{if}~k\in Q_K~\text{else}~0\ .
    \end{align}
\item We rank every QM region $i$ as
    \begin{align}
        a^\mathrm{region}_K = \sum_{k\in Q_K} a^\mathrm{atom}_k
    \end{align}
to quantify how representative the region is for all QM regions.
\item We add the atom indices of the region with the maximum $a^\mathrm{region}_K$ to the final list of QM atom indices $Q$.
\item We set $a_k = 0$ for all $k\in Q$ and repeat steps (ii) to (iv) until $Q$ contains more than $n^\mathrm{region}_\mathrm{target}$ atom indices.
\end{enumerate}

\section{Energy Expression}\label{energy-expr}
We use the cumulant-based energy expression introduced in a previous publication.\cite{k-be} For clarity, we explicitly reproduce the working equations used for energy reconstruction.

\begin{align}
    E &= E_\text{HF} + \sum_{A}^{N_\text{frag}}\sum_{p \in \mathbb{C}_A} \left( \sum_q \left(h_{pq}^A + \frac{1}{2} V_{pq}^{A,[0]} \right) \Delta P_{pq}^A + \frac{1}{2} \sum_{qrs} (pq|rs)^A K_{pqrs}^A\right)\\
    K^A_{pqrs} &= K'^A_{pqrs} + \Delta P^A_{pq} \Delta P_{rs}^A - \frac{1}{2} \Delta P_{pr}^A \Delta P_{sq}^A \\
    &= \Gamma^A_{pqrs} - P^A_{pq} P^A_{rs} + \frac{1}{2}P_{pq}^A P_{sq}^A + \Delta P^A_{pq} \Delta P_{rs}^A - \frac{1}{2} \Delta P_{pr}^A \Delta P_{sq}^A\\
    \Delta P^A &= P^A - P_\text{HF}^A
\end{align}

Here, $E_\text{HF}$ denotes the Hartree-Fock energy of the system. Quantities denoted with superscript $A$ refers to the fragment quantities expressed in the embedding basis. $\mathbb{C}_A$ indicates the center-site indices of the fragment $A$. $h^A$ is the one-electron part of the fragment Hamiltonian, and $V^{A,[0]}$ is the effective Fock potential built using the Hartree-Fock densities. $p, q, r,$ and $s$ are indices over the embedding orbitals, which consist of fragment and bath orbitals. $(pq|rs)$ is the electron repulsion integral in the embedding basis. $P$, $\Gamma$, $K$, and $K'$ refer to the 1-RDM, 2-RDM, approximate cumulant, and true cumulant, respectively.

\section{Results}\label{Results}
We tabulate the full raw data here for clarity. For both systems, we observe that BE$(n)$-in-BE$(m)$ converges to BE$(n)$ for sufficiently large $m$. As we mentioned in the main text, BE$(n)$-in-BE$(n)$ has a notable degradation in its quality with large basis sets and longer chain length. Such errors can be systematically improved by expanding the large basis region as defined with increasing $m$. We have shown in Figure 10 of the main text that this behavior is amplified for BE$(n)$-in-BE$(n)$ and is correlated to the error in electron count. This suggests that the error, in principle, can be corrected using chemical potential matching or some linear correction schemes. To maintain the systematic correctability of our approach, we recommend use of larger $m$ $(> n)$, such as BE$(1)$-in-BE$(2)$ for production-level work.

\subsection{E-Polyacetylene Chains}

\begin{table}[htbp]
\small
\begin{tabular}{l|c|c|clcl|cc|l}
Basis & \begin{tabular}{@{}c@{}}No. \\ C \end{tabular} & 
CCSD &
\begin{tabular}{@{}c@{}}BE$(1)$ \\ -in- \\ BE$(1)$\end{tabular} & \begin{tabular}{@{}c@{}}BE$(1)$ \\ -in- \\ BE$(2)$\end{tabular} & \begin{tabular}{@{}c@{}}BE$(1)$ \\ -in- \\ BE$(3)$\end{tabular} & 
BE$(1)$ &
\begin{tabular}{@{}c@{}}BE$(2)$ \\ -in- \\ BE$(2)$\end{tabular} &
BE$(2)$ &
BE$(3)$ \\ 
\hline
def2-SVP & 4 & -0.5954 & -0.6093 & -0.5523 & -0.5526 & -0.5526 & -0.5976 & -0.5978 & -0.5954\\
 & 8 & -1.1612 & -1.1824 & -1.0352 & -1.0279 & -1.0330 & -1.1732 & -1.1677 & -1.1686\\
 & 12 & -1.7269 & -1.7547 & -1.5177 & -1.5017 & -1.5129 & -1.7502 & -1.7361 & -1.7447\\
 & 16 & -2.2866 & -2.3268 & -2.0000 & -1.9752 & -1.9925 & -2.3272 & -2.3044 & -2.3204\\
 & 20 & -2.8582 & -2.8988 & -2.4823 & -2.4485 & -2.4722 & -2.9042 & -2.8727 & -2.8960\\
\hline
def2-TZVP & 4 & -0.7507 & -0.8319 & -0.5966 & -0.5941 & -0.5941 & -0.7570 & -0.7552 & -0.7507\\
 & 8 & -1.4736 & -1.6420 & -1.1033 & -1.0879 & -1.0878 & -1.5047 & -1.4781 & -1.4845\\
 & 12 & -2.1966 & -2.4511 & -1.6093 & -1.5800 & -1.5812 & -2.2532 & -2.1990 & \\
 & 16 & -2.9197 & -3.2600 & -2.1151 & -2.0716 & -2.0745 & -3.0013 & -2.9198 & \\
 & 20 &  & -4.0687 & -2.6210 & -2.5628 & -2.5677 & -3.7494 & -3.6405 & \\
\hline
def2-QZVP & 4 & -0.8241 & -0.9692 & -0.5992 & -0.5955 & -0.5955 & -0.8339 & -0.8435 & -0.8241\\
 & 8 & -1.6154 & -1.9247 & -1.0931 & -1.0741 & -1.0730 &  &  & \\
 & 12 &  & -2.8788 & -1.5858 & -1.5513 & -1.5503 &  &  & \\
 & 16 &  & -3.8324 & -2.0784 & -2.0279 & -2.0281 &  &  & \\
 & 20 &  & -4.7858 & -2.5709 & -2.5043 & -2.5048 &  &  & \\
\hline \hline
cc-pVDZ & 4 & -0.5949 & -0.6246 & -0.5373 & -0.5365 & -0.5365 & -0.5981 & -0.5976 & -0.5949\\
 & 8 & -1.1589 & -1.2126 & -0.9999 & -0.9898 & -0.9928 & -1.1750 & -1.1656 & -1.1666\\
 & 12 & -1.7227 & -1.7996 & -1.4622 & -1.4417 & -1.4487 & -1.7528 & -1.7318 & -1.7405\\
 & 16 & -2.2866 & -2.3864 & -1.9243 & -1.8933 & -1.9046 & -2.3306 & -2.2979 & -2.3142\\
 & 20 & -2.8504 & -2.9732 & -2.3863 & -2.3447 & -2.3604 & -2.9084 & -2.8641 & -2.8878\\
\hline
cc-pVTZ & 4 & -0.7540 & -0.8343 & -0.6034 & -0.6017 & -0.6017 & -0.7591 & -0.7589 & -0.7540\\
 & 8 & -1.4763 & -1.6373 & -1.1101 & -1.0980 & -1.0988 & -1.5013 & -1.4813 & -1.4877\\
 & 12 & -2.1987 & -2.4393 & -1.6161 & -1.5924 & -1.5957 & -2.2440 & -2.2019 & \\
 & 16 & -2.9211 & -3.2411 & -2.1220 & -2.0862 & -2.0925 & -2.9864 & -2.9223 & \\
 & 20 &  & -4.0427 & -2.6279 & -2.5796 & -2.5893 & -3.7288 & -3.6427 & \\
\hline
cc-pVQZ & 4 & -0.8388 & -0.9792 & -0.6147 & -0.6110 & -0.6110 & -0.8492 & -0.8435 & -0.8388\\
 & 8 & -1.6441 & -1.9417 & -1.1237 & -1.1049 & -1.1036 &  &  & \\
 & 12 & -2.4495 & -2.9029 & -1.6319 & -1.5971 & -1.5959 &  &  & \\
 & 16 &  & -3.8637 & -2.1399 & -2.0886 & -2.0881 &  &  & \\
 & 20 &  & -4.8243 &  &  &  &  &  & \\
\hline
cc-pV5Z & 4 & -0.8727 & -1.0711 & -0.5890 & -0.5843 & -0.5843 &  &  & \\ 
 & 8 &  & -2.1402 & -1.0677 & -1.0447 & -1.0425 &  &  & \\
 & 12 &  & -3.2076 & -1.5452 & -1.5040 & -1.5040 &  &  & \\
 & 16 &  & -4.2743 & -2.0225 &  &  &  &  & \\
 & 20 &  & -5.3409 &  &  &  &  &  & \\
 \end{tabular}
\caption{Correlation energies (Ha) for E-Polyacetylene chains of increasing length.} \label{table:epoly_data}
\end{table}

Table \ref{table:epoly_data} provides the raw data of correlation energies for E-Polyacetylene chains of different length. As noted in the main text, BE$(1)$-in-BE$(m)$ converges to BE$(1)$ with increasing $m$. BE$(2)$-in-BE$(2)$ results closely track the BE$(2)$ and CCSD correlation energies.

\subsection{Polyglycine Chains}

\begin{table}[htbp]
\small
\begin{tabular}{l|c|c|clcl|cc|l}
Basis & \begin{tabular}{@{}c@{}}No. \\ C \end{tabular} & 
CCSD &
\begin{tabular}{@{}c@{}}BE$(1)$ \\ -in- \\ BE$(1)$\end{tabular} & \begin{tabular}{@{}c@{}}BE$(1)$ \\ -in- \\ BE$(2)$\end{tabular} & \begin{tabular}{@{}c@{}}BE$(1)$ \\ -in- \\ BE$(3)$\end{tabular} & 
BE$(1)$ &
\begin{tabular}{@{}c@{}}BE$(2)$ \\ -in- \\ BE$(2)$\end{tabular} &
BE$(2)$ &
BE$(3)$ \\ 
\hline
cc-pVDZ & 2 & -1.4935 & -1.5349 & -1.3805 & -1.3767 & -1.3766 & -1.4987 & -1.4939 & -1.4935\\
 & 4 & -2.7700 & -2.8479 & -2.5250 & -2.5158 & -2.5159 & -2.7821 & -2.7703 & -2.7719\\
 & 6 & -4.0467 & -4.1610 & -3.6695 & -3.6549 & -3.6553 & -4.0657 & -4.0469 & \\
 & 8 &  & -5.4740 & -4.8140 & -4.7940 & -4.7947 & -5.3493 & -5.3236 & \\
 & 12 &  & -8.1001 & -7.1030 & -7.0721 &  & -7.9166 &  & \\
 & 16 &  & -10.7262 & -9.3920 & -9.3503 &  & -10.4839 &  & \\
\hline
cc-pVTZ & 2 & -1.9323 & -2.0772 & -1.6320 & -1.6235 & -1.6235 & -1.9446 & -1.9309 & \\
 & 4 &  & -3.8643 & -2.9608 & -2.9398 & -2.9398 & -3.6152 &  & \\
 & 6 &  & -5.6514 & -4.2896 & -4.2560 &  & -5.2861 &  & \\
 & 8 &  & -7.4386 & -5.6184 & -5.5722 &  & -6.9570 &  & \\
 & 12 &  & -11.0129 & -8.2760 & -8.2046 &  &  &  & \\
 & 16 &  & -14.5872 & -10.9337 & -10.8370 &  &  &  & \\
\hline
cc-pVQZ & 2 &  & -2.4658 & -1.6953 & -1.6826 & -1.6818 &  &  & \\
 & 4 &  & -4.6052 & -3.0575 & -3.0264 & -3.0231 &  &  & \\
 & 6 &  & -6.7447 & -4.4196 & -4.3702 &  &  &  & \\
 & 8 &  & -8.8842 & -5.7818 & -5.7140 &  &  &  & \\
 & 12 &  & -13.1633 & -8.5061 & -8.4016 &  &  &  & \\
 & 16 &  & -17.4423 & -11.2304 &  &  &  &  & \\
\hline
cc-pV5Z & 2 &  & -2.7040 & -1.6377 &  &  &  &  & \\
 & 4 &  & -5.0666 & -2.9353 &  &  &  &  & \\
 & 6 &  & -7.4294 &  &  &  &  &  & \\
 & 8 &  & -9.7921 &  &  &  &  &  & \\
 & 12 &  & -14.5177 &  &  &  &  &  & \\
 & 16 &  & -19.2433 &  &  &  &  &  & \\
\end{tabular}
\caption{Correlation energies (Ha) for Polyglycine chains of increasing length.} \label{table:polyg_data}
\end{table}

Table \ref{table:polyg_data} provides the raw data of correlation energies for Polyglycine chains of different length. We witness a trend similar to that in E-Polyacetylene chains. That is, BE$(1)$-in-BE$(m)$ converges to BE$(1)$ with increasing $m$ and BE$(2)$-in-BE$(2)$ results are close to the BE$(2)$ and CCSD correlation energies. We note that for Polyglycine chains, we have limited number of reference values: the CCSD references for these systems in non-minimal basis sets are expensive calculations. We provide additional data to demonstrate the scope of calculations performed beyond conventional methods.

\bibliography{main}